\pdfoutput=1

\documentclass[journal,12pt,onecolumn,draftclsnofoot,]{IEEEtran}
\normalsize

\ifCLASSINFOpdf
\else
\fi
\normalsize
\usepackage{amsmath,amsfonts}

\usepackage{array}
\usepackage[caption=false,font=normalsize,labelfont=sf,textfont=sf]{subfig}
\usepackage{textcomp}
\usepackage{stfloats}
\usepackage{url}
\usepackage{verbatim}
\usepackage{graphicx}
\usepackage{cite}
\usepackage{multirow}

\usepackage{algpseudocode}
\usepackage[top=2cm, bottom=2cm, left=2cm, right=2cm]{geometry}
\usepackage{color}
\usepackage{mathrsfs}

\usepackage[ruled,linesnumbered]{algorithm2e}

\begin{document}

\title{ Joint Beam Management and SLAM for mmWave Communication Systems
}

\author{
Hang~Que,~%~\IEEEmembership{Staff,~IEEE,}
Jie~Yang,~%\IEEEmembership{Member,~IEEE,} 
Chao-Kai~Wen,~%\IEEEmembership{Senior~Member,~IEEE,} 
Shuqiang~Xia, 
Xiao~Li,~%\IEEEmembership{Member,~IEEE,} 
and~Shi~Jin%,~\IEEEmembership{Senior~Member,~IEEE}
% <-this % stops a space
\thanks{Hang Que, Xiao Li, and Shi Jin are with the National Mobile Communications Research Laboratory, Southeast University, Nanjing, China (e-mail: \{quehang; li\_xiao; jinshi\}@seu.edu.cn). 
Jie Yang is with the School of Automation, Southeast University, Nanjing, China (e-mail: yangjie@seu.edu.cn). 
Jie Yang and Shi Jin are with the Frontiers Science Center for Mobile Information Communication and Security, Southeast University, Nanjing, China. 
Chao-Kai Wen is with the Institute of Communications Engineering, National Sun Yat-sen University, Kaohsiung, 804, Taiwan (e-mail: chaokai.wen@mail.nsysu.edu.tw). 
Shuqiang Xia is with the ZTE Corporation and the State Key Laboratory of Mobile Network and Mobile Multimedia, Shenzhen, China (e-mail: xia.shuqiang@zte.com.cn). 
}
}

%\markboth{IEEE Transactions on Communications}%
%{Submitted paper}

%\markboth{Journal of \LaTeX\ Class Files,~Vol.~001, No.~001, temporary~2022}%


\maketitle
\begin{abstract}
The millimeter-wave (mmWave) communication technology, which employs large-scale antenna arrays, enables inherent sensing capabilities. 
Simultaneous localization and mapping (SLAM) can utilize channel multipath angle estimates to realize integrated sensing and communication design in 6G communication systems. 
However, existing works have ignored the significant overhead required by the mmWave beam management when implementing SLAM with angle estimates. 
This study proposes a joint beam management and SLAM design that utilizes the strong coupling between the radio map and channel multipath for simultaneous beam management, localization, and mapping. 
In this approach, we first propose a hierarchical sweeping and sensing service design. The path angles are estimated in the hierarchical sweeping, enabling angle-based SLAM with the aid of an inertial measurement unit (IMU) to realize sensing service. 
Then, feature-aided tracking is proposed that utilizes prior angle information generated from the radio map and IMU. 
Finally, a switching module is introduced to enable flexible switching between hierarchical sweeping and feature-aided tracking. 
Simulations show that the proposed joint design can achieve sub-meter level localization and mapping accuracy (with an error $<0.5 \;{\rm m}$). Moreover, the beam management overhead can be reduced by approximately 40\% in different wireless environments. 
\end{abstract}


\IEEEpeerreviewmaketitle

%Introduction
\section{Introduction}
\IEEEPARstart{R}{adio} sensing and communication systems are evolving to have similar hardware architecture and deployed spectrum \cite{a0}. 
This trend offers an opportunity to implement the integrated sensing and communication (ISAC) design in 6G millimeter-wave (mmWave) wireless communication systems \cite{a001}. 
ISAC allows wireless communication systems to provide high-quality communication services and sensing services simultaneously, such as high-precision positioning, human gesture recognition, and environmental mapping \cite{a1}.

ISAC research has garnered significant attention from academia and industry in recent years \cite{a2}.
A joint communication-radar system proposed in \cite{a3} achieved symbiotic communication and sensing design by combing a mmWave communication system with a full-duplex radar receiver. 
Strategies for user equipment (UE) positioning, based on Kalman filtering utilizing multipath angle estimates and sensor measurements, were designed in \cite{a4}. In \cite{a5}, a THz imaging prototype with millimeter-level mapping accuracy, similar in size to traditional mobile UEs, was proposed, demonstrating the fine-grained sensing capabilities of 6G communication systems. 
Most existing ISAC designs leverage the similarities between wireless communication and radar systems \cite{a501, a502,a503}, enabling hardware reuse and resulting in benefits in power consumption and spectrum efficiency.

Present studies have introduced the concept of simultaneous localization and mapping (SLAM) to mmWave communication systems in \cite{a6,a601,a602}. 
They proposed SLAM algorithms that utilize the multipath channel state information (CSI) to construct a radio map and locate UE simultaneously. SLAM originally emerged from the field of robotics \cite{a7}, where the robots are required to locate themselves and map their environment through sensor networks, such as inertial measurement unit (IMU), camera \cite{a701}, or light detection and ranging (LiDAR) \cite{a702}. 
With the proliferation of intelligent communication equipment, lightweight sensors are increasingly being equipped on UE, creating a broader space for SLAM research based on various sensors. 
For example, radio map construction is realized through LiDAR SLAM and Wi-Fi sensing in \cite{a8}, where LiDAR measurements generate the radio map of constrained spaces where the Wi-Fi signal cannot reach. 
In \cite{a9}, channel-SLAM based on a particle filter is proposed, which integrates the heading information provided by the IMU to improve UE localization accuracy. 
Existing works have mainly focused on high-accuracy radio map construction and ignored the benefits that SLAM can provide for communication systems. This study aims to establish the association between SLAM and mmWave beam management, which differs from existing works that realize radar-enabled sensing \cite{a501, a502,a503}.

The mobility of UE leads to rapid changes in CSI \cite{a902}, resulting in the changeable beam direction. Therefore, the current release of 3GPP 5G NR adopts a beam management framework consisting of beam sweeping, measurement, determination, and reporting \cite{a10}. 
{\color{black}Angle information of the multipath, such as the angle of arrival (AoA) and angle of departure (AoD), can be extracted during the beam management process using Kalman filters \cite{a1001,a100101}.} %, such as beam tracking and compensation \cite{a1002},
However, reducing the overhead of beam management requires further research \cite{a1003}. For example, the concept of belief propagation is utilized in \cite{a11}, where the beam tracking problem is transformed into a multi-target tracking problem through a probability transfer model that describes the relationship between the historic and current beam management results. 
In \cite{a1101}, a novel extended Kalman filtering framework is proposed for tracking and predicting the kinematic parameters of vehicles, resulting in an overhead-reduced beam tracking scheme. These studies suggest that the key to overhead-reduced beam management is extracting useful information from historical measurements, UE motion status, and wireless environments \cite{a11,a1101,a11011}. 
In our study, we believe that the radio map constructed by SLAM can generate prior angle information that is beneficial for achieving overhead-reduced beam tracking. 

\begin{figure}[]
\centering
\subfloat[]{\hspace{-1mm}
\includegraphics[scale=1]{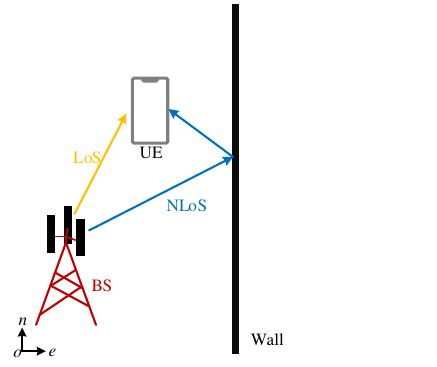}
}
\quad
\subfloat[]{
\includegraphics[scale=1]{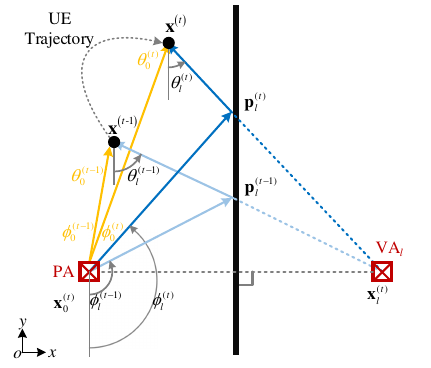}
}
\caption{Illustration of 
(a) the geometric relationship of LoS and NLoS paths between BS and UE and
{\color{black}{(b) the change of the LoS path and the $l$-th NLoS path between time $t-1$ and $t$ with the static VA.}}}
\label{a}%1
\vspace{-8mm}
\end{figure}

Based on the geometric characteristics of mmWave CSI \cite{a901}, the line-of-sight (LoS) path is generated from the base station (BS), whereas the single-bounce specular non-line-of-sight (NLoS) paths can be considered as signals from the mirror images of the BS on different scatter surfaces, as shown in Fig. \ref{a}(a). We regard the BS and the mirror images as the physical anchor (PA) and virtual anchors (VAs), respectively. When the UE is moving, although the channel paths change, the PA and VAs remain static, acting as landmarks that describe the wireless environment \cite{a6,a1102,a1103,a1104,a1105}. Therefore, we can view them as wireless environment features. 
On the one hand, these features simplify the description of the radio map and establish a connection between the multipath and the radio map, providing convenience for feature association in SLAM. On the other hand, these features can be utilized to generate prior information in the angular domain.
Therefore, realizing an ISAC design where the beam management provides angle estimates for SLAM, and SLAM provides prior information for overhead-reduced beam tracking is promising in the mmWave wireless communication system architecture.

However, implementing a joint beam management and SLAM design faces several challenges. 
First, achieving accurate angle estimates through limited beam management overheads is difficult. 
Second, predicting UE motion is necessary for compensating for the rapid shift in beam directions caused by UE mobility. In addition, mmWave signals are easily blocked by obstacles, which may cause the birth or death of channel paths. 
Timely detection of the abrupt change of CSI is required to avoid performance loss. 
Therefore, we develop a joint design composed of beam management and SLAM with the aid of IMU, which can achieve UE localization, radio map construction, and effective beam management simultaneously. The key contributions of this study are as follows: 

\begin{itemize}
{\color{black}
\item \textbf{Beam management and SLAM:} 
The proposed beam management is based on \cite{a12}, which estimates the mmWave channel parameters through measuring beam pairs with different beam-widths. 
We introduce the hierarchical sweeping module, which extracts channel multipath parameters through successive interference cancellation, and add an extra termination criterion to estimate channel multipath numbers. 
SLAM in \cite{a6} can be realized through the path angle estimates. The UE localization and radio map construction results of SLAM are further refined with the aid of IMU, which utilizes historic UE position estimates for position prediction. 
Moreover, we propose a feature-aided tracking module based on the historic radio map from SLAM and the predicted UE position from IMU, extracting multipath parameters with reduced overhead. 

\item \textbf{Switching module:} 
The hierarchical sweeping module can adapt to changeable wireless environments, while the feature-aided tracking module depends on the results from SLAM and IMU and can reduce beam management overhead in stable wireless environments. 
We propose a switching module based on the channel multipath parameters estimated from these two modules to utilize their advantages. 
The switching module enables switch between the hierarchical sweeping and feature-aided tracking modules, completing the joint beam management and SLAM design that can adapt to different wireless environments. 

\item \textbf{Performance analysis:} 
We propose various performance metrics for evaluating both communication and sensing. 
For sensing, we evaluate the angle estimation accuracy, UE localization accuracy, and radio map construction accuracy. 
For communication, we evaluate the channel estimation accuracy, beam management overhead, and spectral efficiency (SE). 
Numerical results show that the proposed joint design can achieve decimeter-level UE localization and radio map construction in different wireless environments. 
Moreover, the overhead-reduced feature-aided tracking module with reliable prior angle information is proved to realize high-accuracy angle estimation. The effectiveness of the proposed switching module is verified.
}
\end{itemize}

The rest of this study is organized as follows. Section II introduces the system model and problem formulation. 
Section III describes the proposed joint design in detail. 
Section IV introduces the performance metrics for sensing and communication. 
Section V presents our experimental results. 
Finally, Section VI concludes the study. 
{\color{black} For clarity, we summarize the main variables in TABLE \ref{table1}.}
{\color{black}
\textbf{Notation---}In this paper, 
${\mathbf A}$ is a matrix, ${\mathbf a}$ is a vector, $a$ is a scalar, and ${\cal A}$ is a set.
$\left( \cdot \right)^{\rm T}$, $\left( \cdot \right)^{\rm H}$, and $\left( \cdot \right)^{\rm -1}$ denote the transpose, conjugate transpose, and matrix inversion operation, respectively. 
For a matrix ${\mathbf A}$, $\left\|{{\mathbf A}}\right\|_F$ is the Frobenius norm, and ${\mathbf A}_{{\cal R},:}$ (${\mathbf A}_{:,{\cal R}}$) are the rows (columns) of the matrix ${\mathbf A}$ with indices in the set ${\cal R}$. %For a matrix ${\mathbf A}$, 
For a vector ${\mathbf a}$, the ${l_2}$ norm is signified by $\left\|{{\mathbf a}}\right\|_2$. 
For a scalar ${ a}$, $\left|{a}\right|$ is the absolute value, and $\left\lceil{a}\right\rceil$ is the smallest integer that is greater than ${ a}$. 
$\mathbb{E}\left[{\cdot}\right]$ indicates the statistical expectation. 
}

\begin{table*}[]
\caption{\color{black}Notations of Important Variables}\vspace{-6mm}
\label{table1}
\vspace{-2mm}
\center
\begin{tabular}{|l|l|l|l|}
\hline
Notation & Definition & Notation & Definition \\ \hline
$t$ & time & $N$ & codebook resolution \\ \hline
${\mathbf{x}^{\left({{t} } \right)}}$ & UE location & ${{r}}_{ab}^j$ & beam measurement signals \\ \hline
${\mathbf{a}^{\left({{t} } \right)}}$ & UE acceleration & ${{{\bf{W}}_{({j,k} )}}}$ & BS sweeping codebook \\ \hline
${\mathbf{v}^{\left({{t} } \right)}}$ & UE velocity & ${{{\bf{F}}_{({j,k} )}}}$ & UE sweeping codebook \\ \hline
${L^{(t)}}$ & channel path number & $\theta _{{\rm{VA}},l}^{(t )},$ $\phi _{{\rm{VA}},l}^{(t )}$ & prior angle values for the $l$-th path \\ \hline
$\mathbf{H}^{(t)}$ & downlink channel & $\theta _{{\rm{scale}},l}^{(t )},$ $\phi _{{\rm{scale}},l}^{(t )}$ & prior angle searching ranges for the $l$-th path \\ \hline
$\mathbf{x}_l^{(t)}$ & PA ($l=0$) or VA ($l \ge 1$) location & ${j^{\rm BS}}$ & initial BS tracking layer index \\ \hline
${\theta _l^{(t)}}$ & AoA of the $l$-th path & $ {j^{\rm UE}}$ & initial UE tracking layer index \\ \hline
${\phi _l^{(t)}}$ & AoD of the $l$-th path & $ {{{\bf{W}}_{({j,{\theta_{\rm VA}}} )}}}$ & BS tracking codebook \\ \hline
${g_l^{(t)}}$ & path gain of the $l$-th path & $ {{{\bf{F}}_{({j,{\theta_{\rm VA}}} )}}}$ & UE tracking codebook \\ \hline
${N_{\rm BS}}$ & BS array antenna number & ${E^{(t )}}$ & channel power \\ \hline
${{N_{\rm UE}}}$ & UE array antenna number & $\Delta _E^{({t,t - 1} )}$ & channel power difference between time $t$ and $t-1$ \\ \hline
${\mathbf{a}_{\rm BS}}\left(\cdot \right)$ & BS array response & ${D^{(t)}} $ & switching symbol \\ \hline
${\mathbf{a}_{\rm UE}}\left(\cdot \right)$ & UE array response & $\hat *$ & estimated $*$, $*$ can be a matrix, a vector, or a scalar \\ \hline
\end{tabular}
\vspace{-4mm}
\end{table*}

%System Model
\vspace{-3.5mm}
\section{System Model}
We consider a mmWave communication system with a static BS and a moving UE, as shown in Fig. \ref{a}. The BS and UE equipped with beamforming architectures perform the beam management periodically to maintain highly directional transmission and extract wireless environment features simultaneously.
\vspace{-3.5mm}
\subsection{Kinetic and Geometric Models}
\vspace{-1.5mm}
The BS and UE are located in the same global two-dimensional (2D) coordinate system with due east as the positive $x$ axis direction and due north as the positive $y$ axis direction, as shown in Fig. \ref{a}(b). The location of UE at time $t$ is denoted as ${\mathbf{x}^{({{t}} )}} = {[ {{x^{({{t}} )}},{y^{(t )}}} ]^{\rm T}}$, and the uniformly accelerated rectilinear motion model is considered to describe the motion of UE during the interval time ${T}$, which is given by
\begin{subequations}
\vspace{-2mm}
\begin{align}
{\mathbf{x}^{\left({t} \right)}} - {\mathbf{x}^{\left({{t} - 1} \right)}} &= {T}{\mathbf{v}^{\left({t - 1} \right)}} + \frac{1}{2}{{T}^2}{\mathbf{a}^{\left({t - 1} \right)}}, \\
{\mathbf{v}^{\left(t \right)}} - {\mathbf{v}^{\left({t - 1} \right)}} &= {T}{\mathbf{a}^{\left({t - 1} \right)}},
\end{align}

\end{subequations}
\noindent where ${\mathbf{v}^{({t} )}},{\mathbf{a}^{({t} )}} \in {\mathbb{R}^{2\times 1}}$ represent the velocity and acceleration vectors of UE in the global 2D coordinate at time $t$, respectively. 

We set the AoA ${\theta _l^{(t)}}$ and AoD ${\phi _l^{(t)}}$ of the $l$-th path ($ l=0,1,..., {L^{(t)}}-1$) at time $t$ start from the negative $y$ axis, as illustrated in Fig. \ref{a}(b).
The PA (BS) is the source of the LoS path ($l=0$), which is located at $\mathbf{x}_0^{(t)}={[ {{x_0^{(t)}},{y_0^{(t)}}} ]^{\rm T}}$. When UE is moving, the LoS path is generated by the static PA, as shown in Fig. \ref{a}(b). 
Thus, ${\theta _0^{(t)}}$ and ${\phi _0^{(t)}}$ satisfy
\vspace{-2mm}
\begin{equation}
{\rm{tan}}\left( \theta _0^{(t)} \right) = {\rm{tan}}\left( \phi _0^{(t)} \right)={\frac{{{y_0^{(t)}} - y^{(t)}}}{{{x_0^{(t)}} - x^{(t)}}}}.
\end{equation}
Meanwhile, the VAs are mirror images of the PA on different scattering surfaces. VAs can be viewed as the source of single-bounce specular NLoS paths ($l\;\ge\;1$). 
The locations of the available VAs at time $t$ are represented by $\mathbf{x}_l^{(t)}={[ {{x_l^{(t)}},{y_l^{(t)}}} ]^{\rm T}}, l=1,2,..., {L^{(t)}}-1$, and the corresponding reflection points are represented by $\mathbf{p}_l^{(t)}={[ {{x_{{\rm p},l}^{(t)}},{y_{{\rm p},l}^{(t)}}} ]^{\rm T}}$. 
Utilizing the geometry relationship in Fig. \ref{a}(b), the NLoS path angles ${\theta _l^{(t)}},{\phi _l^{(t)}}$ with $l\;\ge\;1$ satisfy
\vspace{-2mm}
\begin{subequations}
\label{4}
\begin{align}
{\rm{tan}}\left( \theta _l^{(t)} \right) &= {\frac{{{y^{\left(t \right)}} - y_l^{(t)}}}{{{x^{\left(t \right)}} - x_l^{(t)}}}},\\ 
{\rm{tan}}\left( \phi _l^{(t)} \right)&= {\frac{{{y_0} - x_{{\rm p},l}^{(t)}}}{{{x_0} - y_{{\rm p},l}^{(t)}}}},
\end{align}
\end{subequations}
Therefore, the wireless environment features can be represented by the locations of PA and all possible VAs as $\mathbf{x}_l^{(t)}={[ {{x_l^{(t)}},{y_l^{(t)}}} ]^{\rm T}}, l=0,1,..., {L^{(t)}}-1$. 
\vspace{-3.5mm}
\subsection{Signal and Channel Models}
\vspace{-1.5mm}
For signal model construction, we consider that the BS has $N_{{\rm BS}}$ antennas and one radio frequency (RF) chain with precoder ${\mathbf{f}} \in {\mathbb{C}^{{N_{{\rm BS}}} \times {1}}}$. The UE has $N_{{\rm UE}}$ antennas and one RF chain with combiner ${\mathbf{w}} \in {\mathbb{C}^{{N_{{\rm UE}}} \times {1}}}$. The power constraint is enforced by ${\left\|{\mathbf{f}}\right\|}_2^2={\left\|{\mathbf{w}}\right\|}_2^2=1$.
Thus, when the BS transmitted {\color{black}the signal ${ s}$ restricted to $\left|{ s}\right|=1$}, the received signal ${ r}$ of the UE through the mmWave downlink channel ${\mathbf{H}^{(t)}} \in {\mathbb{C}^{{N_{\rm UE}} \times {{N_{\rm BS}}}}}$ can be represented as
\vspace{-2mm}
\begin{equation}
{ r}= { {\mathbf{w}} ^{\rm H}}{\mathbf{H}^{(t)}} {\mathbf{f}}{ s} + { {\mathbf{w}} ^{\rm{H}}}\mathbf{n},
\end{equation}
where $\mathbf{n} \sim {\mathcal{N}(\mathbf{0},\sigma^2 \mathbf{I})}$ represents the additive Gaussian noise. 
During the beam management process, one beam pair is measured at a time, where $\mathbf{f}$ and $\mathbf{w}$ are the beamforming vectors of BS and UE, respectively. 
The amplitude of the received signal ${r}$ can be used to evaluate the quality of the beam pair $\mathbf{f}$ and $\mathbf{w}$. The qualities of the different beam pairs are measured periodically to monitor CSI change.

For channel model construction, we consider the geometry-based mmWave multipath channel, as shown in Fig. \ref{a}(b). 
The number of the available paths from BS to UE at time $t$ is given by $L^{(t)}$, with the LoS path $(l=0)$ from the BS, and the single-bounce specular NLoS paths $(l=1,2....,L^{(t)}-1)$ from the VAs. 
Thus, the downlink channel $\mathbf{H}^{(t)}$ can be represented as
\vspace{-2mm}
\begin{equation}
\mathbf{H}^{(t)} = \mathop \sum \limits_{l = 0}^{L^{(t)} - 1} {g_l^{(t)}}{\mathbf{a}_{{\rm UE}}}\left({{\theta _l^{(t)}}} \right)\mathbf{a}_{{\rm BS}}^{{\rm H}}\left({{\phi _l^{(t)}}} \right),
\end{equation}
{\color{black} {where ${g_l^{(t)}}$ represents the path gain of the $l$-th path at time $t$, considering the ray tracing path loss model in \cite{a18}.}} 
The antenna array response vectors at UE and BS are represented by ${\mathbf{a}_{\rm UE}}({\theta _l^{(t)}} )$ and ${\mathbf{a}_{\rm BS}}(\phi _l^{(t)} )$ with the $l$-th path's AoA and AoD ${\theta _l^{(t)}},{\phi _l^{(t)}}$ at time $t$. {\color{black} {We consider that the BS and UE are equipped with uniform linear arrays (ULAs). Thus, the array responses can be represented as
\begin{subequations}
\label{5}
\begin{align}
{\mathbf{a}_{\rm UE}}\left({\theta _l^{(t)}} \right)&=\sqrt {\frac{1}{{{N_{{\rm{UE}}}}}}} {\left[ {1,{e^{-j{ \pi}\cos \left({\theta _l^{(t)}}\right)}}, ...,{e^{-j{ \pi}({N_{{\rm{UE}}}} - 1)\cos \left({\theta _l^{(t)}}\right)}}} \right]^{\rm T}}, \\ 
{\mathbf{a}_{\rm BS}}\left({\phi _l^{(t)}} \right)&= \sqrt {\frac{1}{{{N_{{\rm{BS}}}}}}} {\left[ {1,{e^{-j{ \pi}\cos \left({\phi _l^{(t)}}\right)}}, ...,{e^{-j{ \pi}({N_{{\rm{BS}}}} - 1)\cos \left({\phi _l^{(t)}}\right)}}} \right]^{\rm T}},
\end{align}
\end{subequations}
where the angles ${\theta _l^{(t)}},{\phi _l^{(t)}}\in (0,\pi )$\footnote{\color{black}The channel paths with AoA and AoD beyond $(0,\pi)$ can be distinguished by multiple ULAs with different orientations equipped on UE and BS.}.}}

\vspace{-3.5mm}
\subsection{Problem Formulation}
\vspace{-1.5mm}
We consider an indoor scenario where the UE enters an unknown environment served by a single BS. 
GPS signals are unavailable indoors, making high-accuracy localization of a moving UE difficult to achieve. 
SLAM can be introduced to estimate the locations of UE, PA, and VAs. However, SLAM requires continuous environment measurements from the UE, and measurement errors can result in performance loss. 
%On the other hand, mmWave channel paths are easily blocked by surrounding obstacles, and the CSI changes rapidly. Periodical beam management can be introduced to search for beam pairs with high beamforming gain. 
Additionally, mmWave channel paths are susceptible to blockage by surrounding obstacles, resulting in rapid changes in CSI. To overcome this, periodic beam management can be introduced to search for beam pairs with high beamforming gain. 
%However, the beam sweeping procedure invokes numerous beam pairs between the UE and BS, leading to large overhead consumption. 
However, the beam sweeping procedure invokes testing numerous beam pairs between the UE and BS, resulting in significant overhead consumption. 
{\color{black}{To address these challenges, this study proposes a joint design where SLAM and beam management can mutually assist each other. By leveraging this joint design, the UE can achieve high-accuracy localization, construct a radio map of the environment, and effectively manage beams. }}

% Algorithm Design

\begin{figure*}
\centering
\includegraphics[scale=0.466]{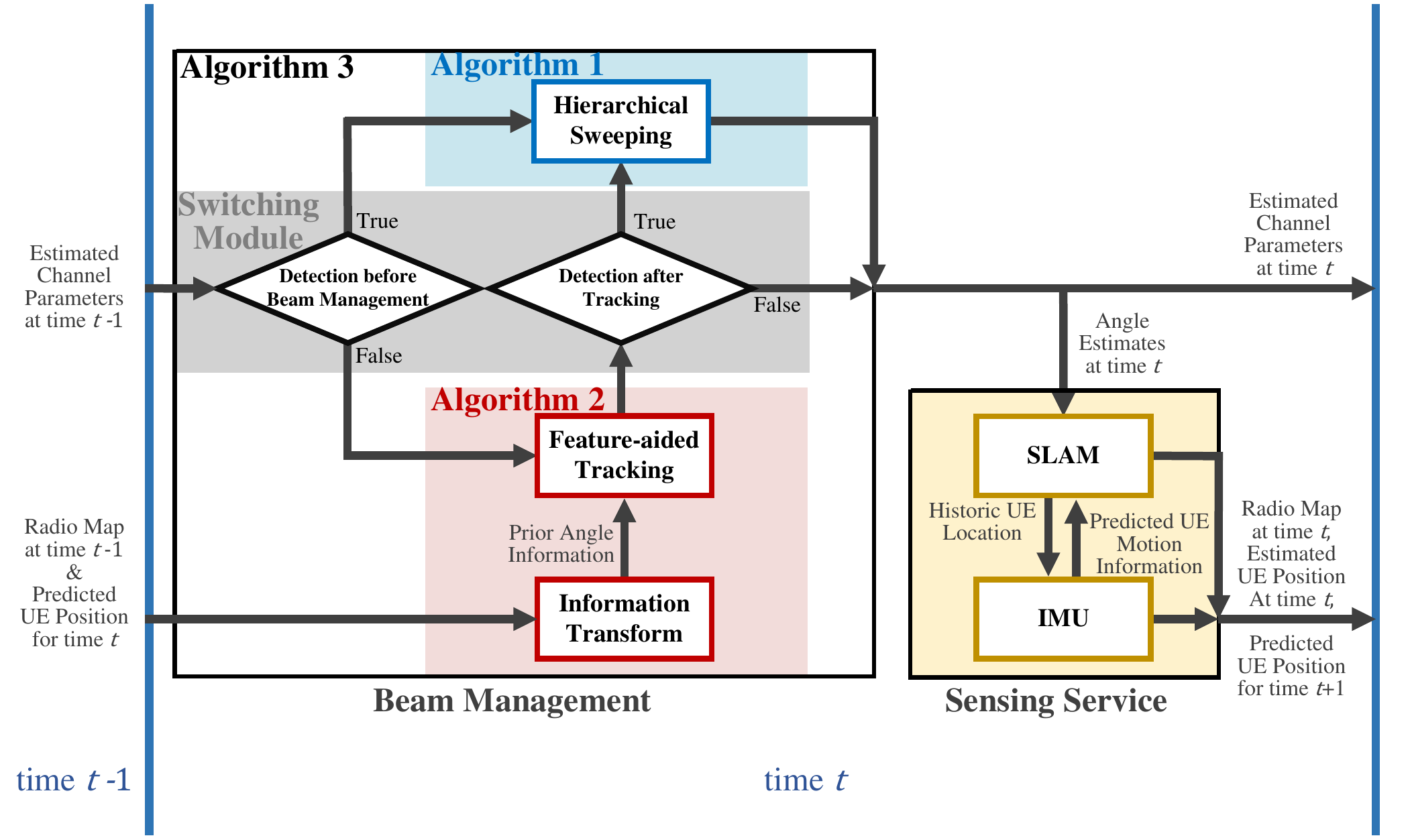}\vspace{-2.5mm}
\caption{{\color{black}{Block diagram of the proposed joint beam management and SLAM design.}}}
\label{c}%3
\vspace{-6mm}
\end{figure*}
\vspace{-2.5mm}
\section{Joint Beam Management and SLAM}
The proposed joint design consists of four modules in different colored regions, as shown in Fig. \ref{c}. 
The blue region represents the hierarchical sweeping module, which covers the full-angle region.
The red region represents the feature-aided tracking module with reduced overhead consumption. 
The gray region represents the switching module that controls the choice of the hierarchical sweeping and feature-aided tracking modules. 
The yellow region consists of SLAM and IMU, assisting each other to realize UE localization, radio map construction, and UE position prediction. 
The hierarchical sweeping, feature-aided tracking, and switching modules complete the beam management, while SLAM and IMU provide sensing services. 
At the end of time $t$, the proposed joint design can estimate the current channel parameters, the UE position, the radio map, and predict the UE position for time $t+1$.

Fig. \ref{c} shows the working procedure at time $t$. First, the switching module utilizes the channel parameters estimated at time $t-1$ to choose between the hierarchical sweeping and feature-aided tracking modules. 
The feature-aided tracking module is based on the historic radio map and predicted UE position. The effectiveness of the estimates from the feature-aided tracking module is determined by the switching module.
Then, the path angle estimates of the hierarchical sweeping or feature-aided tracking module are utilized to realize the sensing services. 
Finally, the results of the beam management and sensing service are transmitted to time $t+1$, marking the end of the joint design at time $t$. The implementation of different modules depends on the results of the others. 
Notably, as the UE has no prior information at first, the hierarchical sweeping module is performed for initialization at $t=1$. 
Therefore, we first introduce the hierarchical sweeping module. 
Then, the sensing service based on path angle estimates is introduced, followed by the feature-aided tracking module utilizing the results of the sensing service. 
Finally, the switching module is introduced to complete the joint design.

\vspace{-3.5mm}
\subsection{Hierarchical Sweeping Module}
\begin{figure*}
\centering
\centering
\subfloat[]{\hspace{-4mm}
\includegraphics[scale=0.68,trim=0 0 0 0,clip]{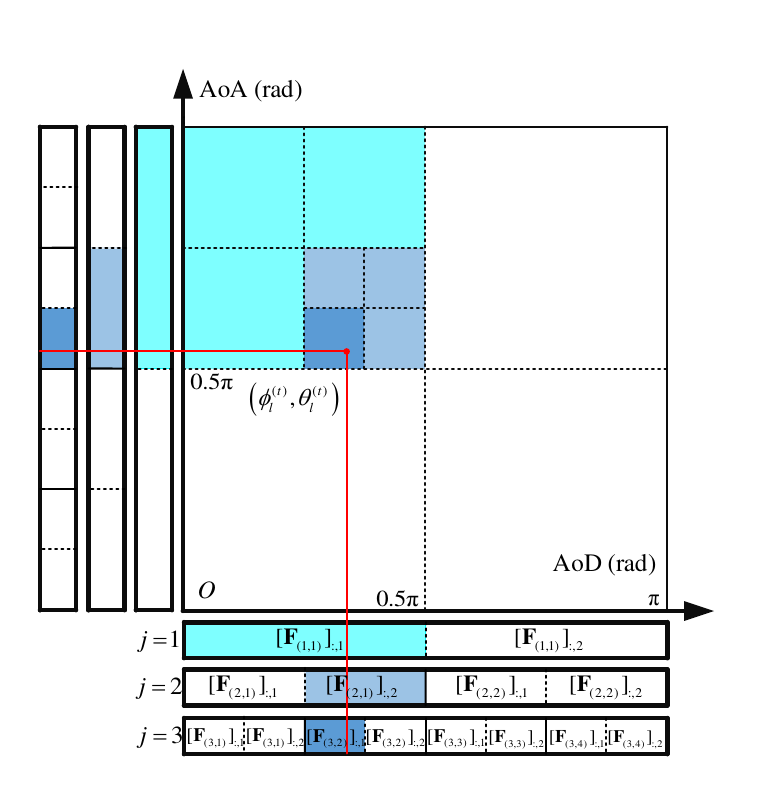}
}
\quad
\subfloat[]{\hspace{-4mm}
\includegraphics[scale=0.68,trim=0 0 0 0,clip]{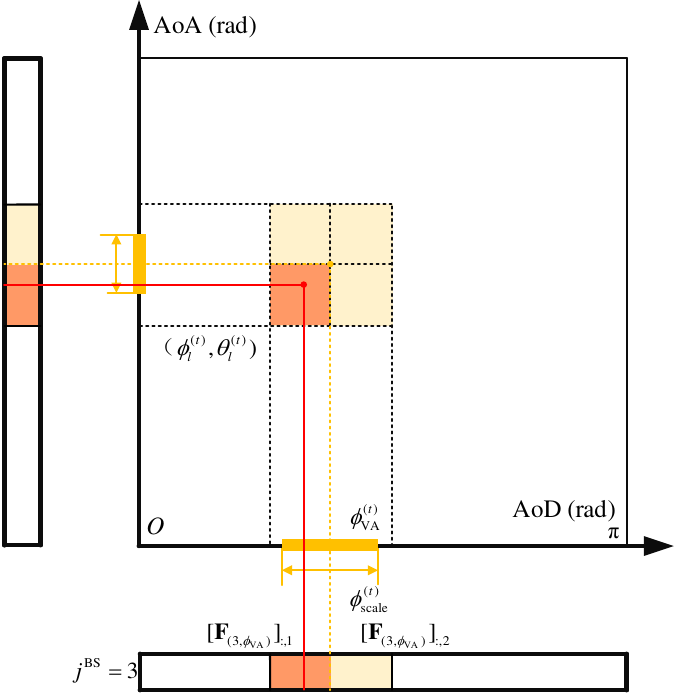}
}
\caption{Illustration of (a) hierarchical sweeping codebook and (b) feature-aided tracking codebook for the $l$-th path.}
\label{d}%4
\vspace{-6mm}
\end{figure*}

\vspace{-1.5mm}
{\color{black}
The hierarchical sweeping module considers the hierarchical codebook proposed in \cite{a12}, which consists of beam pairs with different beam-widths, as shown in Fig. \ref{d}(a). 
The axes in Fig. \ref{d} describe the AoD and AoA of channel paths. For example, the red mark with coordinate $(\phi_{l}^{(t)},\theta_{l}^{(t)})$ represents the $l$-th path with AoD $\phi_{l}^{(t)}$ and AoA $\theta_{l}^{(t)}$. 
The codebooks under the AoD and AoA axes are equipped by BS and UE, respectively. 
For example, the codebook of BS consists of $J$ layers, and the $j$-th layer consists of $2^{j-1}$ subsets. 
The $k$-th $(k=1,...,2^{j-1})$ subset in the $j$-th $(j=1,...,J)$ layer ${ {{{\bf{F}}_{({j,k} )}}} } \in {\mathbb{C}^{{N_{{\rm BS}}} \times {2}}}$ consists two different beamforming vectors.
}

%The sweeping procedure of the $l$-th path is taken as an example.
The generation of ${ {{{\bf{F}}_{({j,k} )}}} }$ is shown as follows. First, the discrete AoD and AoA are set as ${\bar \phi _u} = \frac{{\pi u}}{N},{\bar \theta _v} = \frac{{\pi v}}{N},{\rm{\;}}u,v \in \{ {1, ...,N } \}$, where $\{ {1, ...,N} \}$ represents the discrete angle domain, and ${\rm{\pi }}/N$ represents the corresponding angular resolution.
One subset ${ {{{\bf{F}}_{({j,k} )}}} }$ is selected in one layer, and two vectors are measured during the sweeping procedure. For example, in the third layer of Fig. \ref{d}(a), the second subset is selected. 
Thus, the beamforming vectors ${[ {{{\bf{F}}_{({3,2} )}}} ]_{:,1}}$ and ${[ {{{\bf{F}}_{({3,2} )}}} ]_{:,2}}$ are measured. 
The design of the $m$-th $(m=1,2)$ beamforming vector of the $k$-th $(k=1,...,2^{j-1})$ subset in the $j$-th $(j=1,...,J)$ layer 
${[ {{{\bf{F}}_{({j,k} )}}} ]_{:,m}} \in {\mathbb{C}^{{N_{{\rm BS}}} \times {1}}}$ satisfies the antenna pattern equation sets
\begin{equation}
\label{7}
\left[ {{{\bf{F}}_{\left({j,k} \right)}}} \right]_{:,m}^{\rm H}{{\bf{a}}_{\rm BS}}\left({{{\bar \phi }_u}} \right) = \left\{ {\begin{array}{*{20}{c}}
{{C_j},{\rm{\;if\;}}u \in {{\cal I}_{\left({j,k,m} \right)}}},\\
{0,{\rm{\;\;\;if\;}}u \notin {{\cal I}_{\left({j,k,m} \right)}}},
\end{array}} \right.
\end{equation}
\noindent where the set ${{\cal I}_{({j,k,m} )}} = \{ \frac{N}{{{2^j}}}({2({k - 1} ) + m - 1} ) + 1, ...,
\frac{N}{{{2^j}}} ({2({k - 1} ) + m} ) \}$ describes the projection range of ${[ {{{\bf{F}}_{({j,k} )}}} ]_{:,m}}$ over the discrete angle domain. 
For example, in the second layer of Fig. \ref{d}(a), the beamforming vector ${[ {{{\bf{F}}_{({2,1} )}}} ]_{:,2}}$ covers the discrete angle domain ${{\cal I}_{({2,1,2} )}} =\{\frac{N}{4}+1, \frac{N}{4}+2,..., \frac{N}{2} \}$ corresponding to the continuous angle domain $[ {(\frac{N}{4}+1)\frac{\pi}{N}, \frac{\pi}{2} }]$.
The normalization constant ${C_j}$ is set for power constraint ${\| {{{\bf{F}}_{({j,k} )}}} \|_F} = 1$. The approximate solution of (\ref{7}) is given by \cite{a12}
\begin{equation}
\label{8}
{\left[ {{{\bf{F}}_{\left({j,k} \right)}}} \right]_{:,m}} = {C_j}{\left({{{\bf{A}}_{{\rm{BS,D}}}}{\bf{A}}_{{\rm{BS,D}}}^{\rm H}} \right)^{ - 1}}{{\bf{A}}_{{\rm{BS,D}}}}{\left[ {{{\bf{G}}_{\left({j,k} \right)}}} \right]_{:,m}},
\end{equation}
where ${{\bf{A}}_{{\rm{BS,D}}}} = [ {{{\bf{a}}_{{\rm{BS}}}}({{{\bar \phi }_0}} ),{{\bf{a}}_{{\rm{BS}}}}({{{\bar \phi }_1}} ), ...,{{\bf{a}}_{{\rm{BS}}}}({{{\bar \phi }_{N - 1}}} )} ]$, and ${{\bf{G}}_{({j,k} )}} \in {\mathbb{R}^{N \times {2}}}$ is a matrix with the elements of the $m$-th column and the $u$-th row satisfying $u \in {{\cal I}_{({j,k,m} )}}$ set as 1 whereas others set as 0. 
{\color{black}{Additionally, the codebook $ {{{\bf{W}}_{({j,k} )}}}\in {\mathbb{C}^{{N_{{\rm UE}}} \times {2}}} $ at the UE side can be generated similarly.}}

%The codebook design can be promoted to the UE as $ {{{\bf{W}}_{({j,k} )}}}\in {\mathbb{C}^{{N_{{\rm UE}}} \times {2}}} $.

{\color{black}{We propose a sweeping scheme that utilizes the hierarchical codebook design to estimate the multipath channel parameters path by path. Additionally, the scheme incorporates the successive interference cancellation technique \cite{a1106} to mitigate interference from previously estimated paths.}} 
Before the sweeping procedure for the $l$-th ($l \ge 2$) path, the sweeping procedures for the first $l-1$ paths have already been completed, and the corresponding results can be expressed as $\hat \theta _{l'}^{(t )},\hat \phi _{l'}^{(t )},\;\hat g_{l'}^{(t )}$ for $l' = 0, ...,l - 2$. 
Thus, the partial channel estimation ${{\bf{\hat H}}}_{{\rm{part}},l-1}$ can be calculated as\footnote{{\color{black}{For the $1$-st path ($l = 1$) case, no sweeping procedure has been performed yet, and the partial channel estimation is initialized to ${\bf 0}$.}}}

{\color{black}
\begin{equation}\label{9}
%{{\bf{\hat H}}}_{{\rm{part}},l-1}={ \mathop \sum \limits_{{l'}= 0}^{l - 2} \hat g_{l'}^{\left(t \right)}{{\bf{a}}_{{\rm{UE}}}}\left({\hat \theta _{l'}^{\left(t \right)}} \right){\bf{a}}_{{\rm{BS}}}^{\rm H}\left({\hat \phi _{l'}^{\left(t \right)}} \right)}.
{{\bf{\hat H}}}_{{\rm{part}},l-1} = \left\{ {\begin{array}{*{20}{l}}
{{\bf{0}},\;\;\;\;\;\;\;\;\;\;\;\;\;\;\;\;\;\;\;\;\;\;\;\;\;\;\;\;\;\;\;\;\;\;\;\;\;\;\;\;\;{\rm{if}}\; l=1},\\
{ \mathop \sum \limits_{{l'}= 0}^{l - 2} \hat g_{l'}^{\left(t \right)}{{\bf{a}}_{{\rm{UE}}}}\left({\hat \theta _{l'}^{\left(t \right)}} \right){\bf{a}}_{{\rm{BS}}}^{\rm H}\left({\hat \phi _{l'}^{\left(t \right)}} \right),\;{\rm{if}} \; l\ge2}.
\end{array}} \right. 
\end{equation}
}\noindent The $l$-th sweeping procedure starts from the first layer and ends at the $J$-th layer. 
For the $j$-th layer, as UE and BS select one subset consisting of two beamforming vectors, a total of four different beam pairs are measured.
The corresponding received signals ${{r}}_{ab}^j$ with $a,b\in \{1,2\}$ can be represented as
\begin{equation}
{{r}}_{ab}^j = \left[ {{{\bf{W}}_{\left({j,{k_{{\rm{UE}}}^j}} \right)}}} \right]_{:,a}^{\rm H}\left({{\bf{H}}^{\left(t \right)}{{\left[ {{{\bf{F}}_{\left({j,{k_{{\rm{BS}}}^j}} \right)}}} \right]}_{:,b}}{{s}} + {\bf{n}}} \right),
\end{equation}
\noindent where ${k_{{\rm{UE}}}^j},{k_{{\rm{BS}}}^j}$ are the indexes of the beamforming subset selected by UE and BS for the $j$-th layer. 
Based on (\ref{9}), the residual of the received signal ${{r}}_{{\rm r},ab}^j( l ) $ can be given by
\begin{equation}\label{11}
{{r}}_{{\rm r},ab}^j\left( l \right) = {{r}}_{ab}^j-\left[ {{{\bf{W}}_{\left({j,{k_{{\rm{UE}}}^j}} \right)}}} \right]_{:,a}^{\rm H}
{{\bf{\hat H}}}_{{\rm{part}},l-1}
{{\left[ {{{\bf{F}}_{\left({j,{k_{{\rm{BS}}}^j}} \right)}}} \right]}_{:,b}},
\end{equation}
Then, we set ${m_{{\rm{UE}}}^{j}},{m_{{\rm{BS}}}^{j}}\in \{1,2\}$ as the beamforming vector index of UE and BS corresponding to the maximal ${{r}}_{{\rm r},ab}^j( l ) $, satisfying 
$( {m_{{\rm{UE}}}^j,m_{{\rm{BS}}}^j} ) = \mathop {{\rm{arg\;max}}}| {{r}}_{{\rm r},m_{{\rm{UE}}}^jm_{{\rm{BS}}}^j}^j (l)|$.
After that, the sweeping procedure moves from the $j$-th layer to the $(j+1)$-th one. The corresponding indexes ${k_{{\rm{UE}}}^{j+1}},{k_{{\rm{BS}}}^{j+1}}$ are updated by
\begin{subequations}
\begin{align}
k_{{\rm{UE}}}^{j + 1} &= k_{{\rm{UE}}}^j + \left({m_{{\rm{UE}}}^j - 1} \right),\\
k_{{\rm{BS}}}^{j + 1} &= k_{{\rm{BS}}}^j + \left({m_{{\rm{BS}}}^j - 1} \right),
\end{align}
\end{subequations}
\noindent The sweeping procedure repeats from $j=1$ to $J$, reducing the AoA and AoD range gradually, as shown in Fig \ref{d}(a). 
Finally, at the $J$-th layer, the estimated channel parameters of the $l$-th path $\hat \theta _{l}^{(t )},\;\hat \phi _{l}^{(t )},\;\hat g_{l}^{(t )}$ can be represented by
\begin{subequations}\label{12}
\begin{align}
\hat \theta _{l}^{\left(t \right)} = &\frac{{\pi \left({k_{\rm UE}^J + m_{\rm UE}^J}+0.5 \right)}}{N},\\
\hat \phi _{l}^{\left(t \right)} = &\frac{{\pi \left({k_{\rm BS}^J + m_{\rm BS}^J}+0.5 \right)}}{N},\\
\hat g_{l}^{\left(t \right)} =& {{r}}_{{\rm r},m_{{\rm{UE}}}^Jm_{{\rm{BS}}}^J}^J\left( l \right).
\end{align}
\end{subequations}
The estimated path gain $\hat g_{l}^{(t )}$ depends on the measurement through the beam pair 
selected in the $J$-th layer. We consider simply repeating and averaging such measurements for adequate times to refine $\hat g_{l}^{(t )}$.
Thus, channel parameters of the $l$-th path are estimated through the above-mentioned sweeping procedure. The parameters of the remaining paths can be estimated through similar procedures.

In addition, by adding a termination criterion after the first layer's measurement of every single path, the path number ${L^{(t )}}$ can be estimated from the difference of the received signal. 
For example, after the measurement in the first layer of the $l$-th path's sweeping procedure, we can get the residuals ${{r}}_{{\rm r},ab}^1(l),\; a,b\in \{1,2\}$ based on (\ref{11}). 
When all channel paths are considered in (\ref{9}), the residuals should be small and have a little difference in amplitude.
We set the residual threshold and amplitude difference threshold as ${ r}_{\rm min}$ and ${{\rm{\Delta }}^{{\rm{min}}}_{r}}$, respectively.
Thus, when $\mathop {\max }\limits_{a,b = 1,2} | {{r}}_{{\rm r},ab}^1(l)| < {{{r }}_{{\rm{min}}}}$ and $\mathop {\max }\limits_{a,b,c,d = 1,2} | {{{r}}_{{\rm r},ab}^1(l) - {{r}}_{{\rm r},cd}^1(l)} | < {{\rm{\Delta }}_{{\rm{min}}}}$ are satisfied, the hierarchical sweeping module at time $t$ is terminated, and the estimated channel parameters can be represented as
$\hat \theta _{l'}^{(t )},\hat \phi _{l'}^{(t )},\hat g_{l'}^{(t )},l'= 0,1, ...,{\hat L}^{(t )} - 1$
with ${\hat L}^{(t )}=l$. Otherwise, the $(l+1)$-th path's sweeping procedure continues. {\color{black} Algorithm 1 summarizes the proposed hierarchical sweeping module.}

\begin{algorithm}[t]
\caption{{\color{black}Hierarchical Sweeping}}\label{A1}
\KwIn{Residual threshold ${ r}_{\rm min}$; Amplitude difference threshold ${{\rm{\Delta }}^{{\rm{min}}}_{r}}$;}
Set $l = 1$\;
\While{$l \leq L_{d}$}{
Calculate partial channel estimation {\color{black}using} {(9)}\;
Set $j = 1$\;
\For{$j \leq J$}
{
Run beam measurement {\color{black}using} {(10)} with beamforming vector {\color{black}generated by} {(8)}\;
Calculate residual signal {\color{black}using} {(11)} filtering partial channel estimation {\color{black}using} {(9)}\;
\If{$j= 1$, $\mathop {\max }\limits_{a,b = 1,2} | {{r}}_{{\rm r},ab}^1(l)| < {{{r }}_{{\rm{min}}}}, {\rm{and}} \mathop {\max }\limits_{a,b,c,d = 1,2} | {{{r}}_{{\rm r},ab}^1(l) - {{r}}_{{\rm r},cd}^1(l)} | < {{\rm{\Delta }}_{{\rm{min}}}}$}{
\textbf{end while}\\
}
Calculate the subset index {\color{black}using} {(12)} for the $\left(j+1\right)$-th layer\;
Set {$j= j+1$}\;
}
Run channel parameters estimation {\color{black}using} {(13)}\;
Set {$l= l+1$}\;
}
Set {${\hat L}^{(t )}= l$}\;
\KwOut{channel parameter estimation $\hat \theta _{l'}^{(t )},\hat \phi _{l'}^{(t )},\hat g_{l'}^{(t )}$ for $l'= 0,1, ...,{\hat L}^{(t )} - 1$;}

\end{algorithm}

\vspace{-2.5mm}
\subsection{Sensing Service}
For realizing sensing services, IMU utilizes the historic UE position estimated in SLAM for position prediction, whereas SLAM utilizes the angle estimates from beam management and the measurements from IMU for localization and mapping.

The output of IMU at time $t$ can be represented as the linear acceleration ${{\bf{\hat a}}^{(t )}} = {[ {{{\hat a}}_x^{({t} )},{{\hat a}}_y^{({t} )}} ]^{\rm T}}$ under the global coordinate system. We assume that UE is in a fixed motion pattern, and ${{\bf{\hat a}}^{(t )}}$ can be modeled as
\begin{equation}
{{\bf{\hat a}}^{\left(t \right)}} = {{\bf{a}}^{\left(t \right)}} + {{\bf{n}}_{{\rm{IMU}}}},
\end{equation}
\noindent where ${{\bf{n}}_{{\rm{IMU}}}} \sim {\mathcal{N}(\mathbf{0},\sigma _{{\rm{IMU}}}^2 \mathbf{I})}$ represents the additive Gaussian noise for IMU. The estimated UE trajectory ${{\bf{\hat x}}^{(t )}} = {[ {{{\hat x}^{({t} )}},{{\hat y}^{(t )}}} ]^{\rm T}}$ exported by SLAM (which is introduced in (\ref{15})) is considered to refine the IMU measurements for reducing the integrating error. For example, the predicted UE location ${\bf{\hat x}}{'^{(t )}}$ and velocity ${{\bf{\hat v}}'^{(t )}}$ for time $t$ can be represented as
\begin{subequations}\label{14}
\begin{align}
{\bf{\hat x}}{'^{\left(t \right)}} =&{{{\bf{\hat x}}}^{\left({t - 1} \right)}} + { T}{{{\bf{\hat v}}}^{\left({t - 1} \right)}} + \frac{{{ T}^2}}{2}{{{\bf{\hat a}}}^{\left({t - 1} \right)}},\\
{{{\bf{\hat v}}}'^{\left(t \right)}} =& {{{\bf{\hat v}}}^{\left({t - 1} \right)}} + { T}{{{\bf{\hat a}}}^{\left({t - 1} \right)}},
\end{align}
\end{subequations}
where ${\bf{\hat v}}{^{({t - 1} )}} =\frac{1}{ T}({{{{\bf{\hat x}}}^{({t - 1} )}} - {{{\bf{\hat x}}}^{({t - 2} )}}} ) + \frac{T}{2}{{{\bf{\hat a}}}^{({t - 2} )}}$ is the refined velocity estimate at time $t-1$ utilizing the SLAM results. The predicted results ${\bf{\hat x}}{'^{(t )}},{{\bf{\hat v}}'^{(t )}}$ only depend on the historical results at times $t-1$ and $t-2$, reducing the integrating error of IMU drastically.

The SLAM algorithm proposed in \cite{a6} is introduced to estimate the states and locations of the UE, PA, and VAs.
First, the input of SLAM in \cite{a6} at time $t$ includes the angle estimates $\hat \theta _l^{(t )},\hat \phi _l^{(t )},l = 0,1, ...,{\hat L}^{(t )} - 1$ from the beam management. 
Second, the UE state prediction (position ${\bf{\hat x}}{'^{(t )}}$ and velocity ${{\bf{\hat v}}'^{(t )}}$) in \cite{a6} is based on a linear, near constant-velocity motion model in \cite{a19}. In this study, the UE state can be predicted directly by IMU through (\ref{14}). 
Finally, the particle-based implementation of the SLAM in \cite{a6} is considered. 
Let ${\bf{\hat x}}_{{\rm UE},p}^{(t )}={{{[ {\hat x_{{\rm UE},p}^{(t )},\hat y_{{\rm UE},p}^{(t )}} ]}^{\rm T}}},{\bf{\hat x}}_{l,p}^{(t )}={{{[ {\hat x_{l,p}^{(t )},\hat y_{l,p}^{(t )}} ]}^{\rm T}}},p=1,...,P$ represent the $p$-th particle of the estimated UE and the $l$-th anchor with the total particle number $P$. 
Thus, the estimated UE and anchor positions ${{\bf{\hat x}}^{(t)}}= {[ {{{\hat x}^{({t} )}},{{\hat y}^{(t )}}} ]^{\rm T}},{\bf{\hat x}}_l^{(t )}={[ {\hat x_l^{(t )},\hat y_l^{(t )}} ]^{\rm T}}$ are in the respective centers of their particles, given by
\begin{subequations}
\begin{align}\label{15}
{{\bf{\hat x}}^{\left(t \right)}}& = \frac{1}{P}\sum\limits_{p = 1}^P{\bf{\hat x}}_{{\rm UE},p}^{\left(t \right)},\\
{\bf{\hat x}}_l^{\left(t \right)}& = \frac{1}{P}\sum\limits_{p = 1}^P {\bf{\hat x}}_{l,p}^{\left(t \right)},\;l = 0,1, ..., {\hat L}^{\left(t \right)} - 1.
\end{align}
\end{subequations}

\vspace{-2.5mm}
\subsection{Feature-aided Tracking Module}
The feature-aided tracking module utilizes historical results from the sensing service, as shown in the red region of Fig. \ref{c}. 
First, the information transform sub-module generates prior angle information. Then, the feature-aided tracking sub-module realizes overhead-reduced beam tracking. 
\subsubsection{Information Transform}
The prior angle information at time $t$ includes the prior angle values $\theta _{{\rm{VA}},l}^{(t )},\phi _{{\rm{VA}},l}^{(t )}\in (0,\pi )$ and the prior angle searching ranges $\theta _{{\rm{scale}},l}^{(t )},\phi _{{\rm{scale}},l}^{(t )}\in (0,\pi )$ for ${\hat L}^{({t - 1} )}$ paths. 
The true angles are believed to be in the ranges centered on $\theta _{{\rm{VA}},l}^{(t )}$ (or $\phi _{{\rm{VA}},l}^{(t )}$) with widths equal to $\theta _{{\rm{scale}},l}^{(t )}$ (or $\phi _{{\rm{scale}},l}^{(t )}$).
The prior angle information is generated through the predicted UE location ${\bf{\hat x}}{'^{(t )}}$ from IMU and the radio map ${\bf{\hat x}}_l^{({t - 1} )},{\bf{\hat x}}_{l,p}^{(t-1 )},l = 0,1, ...,{\hat L}^{({t - 1} )} - 1$ from SLAM. The LoS and NLoS paths are discussed separately because of their different characteristics.

\begin{figure*}
\centering
\subfloat[]{
\includegraphics[scale=0.7]{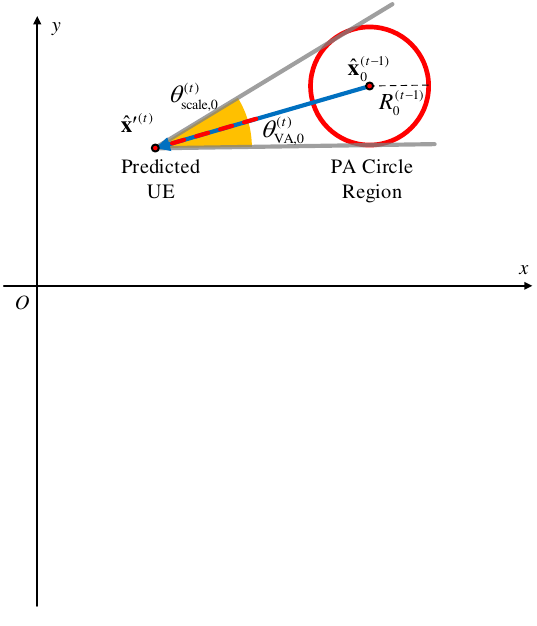}
}
\quad
\subfloat[]{
\includegraphics[scale=0.7]{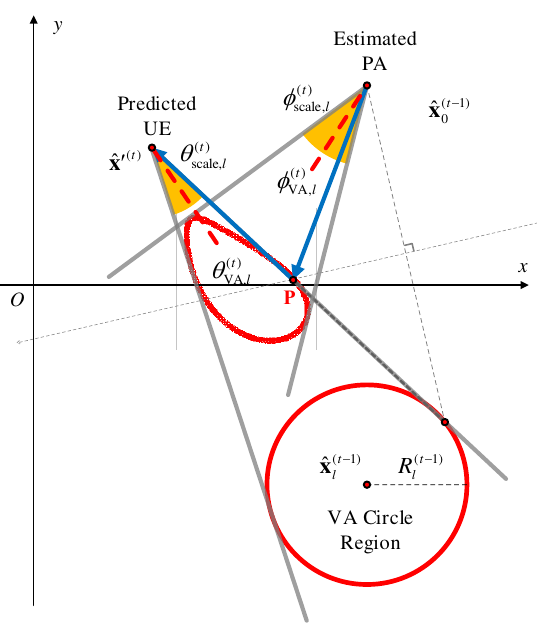}
}
\caption{Illustration of generating prior angle information from the historic radio map and predicted UE location for (a) the LoS path and (b) the NLoS path.}
\label{e}%5
\vspace{-6mm}
\end{figure*}

For the LoS path $(l=0)$, the prior angle values $\theta _{{\rm{VA,0}}}^{(t )}$ and $\phi _{{\rm{VA,0}}}^{(t)}$ are viewed as AoA and AoD of the path from estimated PA ${\bf{\hat x}}_0^{({t - 1} )}$ to predicted UE ${\bf{\hat x}}{'^{(t )}}$, as illustrated in Fig. \ref{e}(a). Thus, $\theta _{{\rm{VA,0}}}^{(t )},\phi _{{\rm{VA,0}}}^{(t )}$ can be calculated as
\begin{subequations}
\begin{align}
\theta _{{\rm{VA,0}}}^{\left(t \right)} &= {\rm{atan}}\left({\frac{{\hat y_0^{\left({t - 1} \right)} - \hat y{'^{\left(t \right)}}}}{{\hat x_0^{\left({t - 1} \right)} - \hat x{'^{\left(t \right)}}}}} \right)+0.5\pi,\label{17a}\\
\phi _{{\rm{VA,0}}}^{\left(t \right)} &= \theta _{{\rm{VA,0}}}^{\left(t \right)} \pm \pi. \label{17b}
\end{align}
\end{subequations}
The prior angle searching ranges are calculated to cover most of the PA/VA particles. 
As the coordinate particle $\hat x_{l,p}^{(t )} $ and $\hat y_{l,p}^{(t )}$ can be approximated as Gaussian distribution, the distance between the estimated PA/VA and the particles ${R^{(t-1)}_{l,p}}=\|{\bf{\hat x}}_l^{(t )}-{\bf{\hat x}}_{l,p}^{(t )}\|_2$ obeys the Rayleigh distribution.
Thus, the statistical mean $\overline R _{l,p}^{(t-1 )}$ and variance $\sigma^2_{{R^{(t-1 )}_{l,p}}}$ can be calculated as
\begin{align}
\overline R _{l,p}^{\left(t-1 \right)}&= \frac{1}{P}\sum\limits_{p = 1}^P{R^{\left(t-1 \right)}_{l,p}},\\
\sigma^2_{{R^{\left(t-1 \right)}_{l,p}}} &=\frac{1}{P}\sum\limits_{p = 1}^P{{\left( {R^{\left(t-1 \right)}_{l,p}} - \overline R _{l,p}^{\left(t-1 \right)} \right)}^{2}}.
\end{align}
Then, the PA/VA circular region ${\bf{C}}^{(t-1)}_l$ centered on ${\bf{\hat x}}_l^{({t - 1})}$ and with radius ${R^{(t-1)}_{l}}={{\rm C}_{\rm angle}}\sigma_{{R^{(t-1)}_{l,p}}}$ is set for LoS and NLoS paths with the angle searching range constant ${{\rm C}_{\rm angle}}$, as illustrated in Fig. \ref{e}(a) and (b), respectively. 
With the accurate UE and BS location, the circular regions with ${{\rm C}_{\rm angle}}\ge3.2759$ can cover all PA/VA particles with over 90\% confidence according to the characteristics of Gaussian distribution. 
Then, the prior angle searching range of the LoS path $\theta _{{\rm{scale,0}}}^{(t )}$ and $\phi _{{\rm{scale,0}}}^{(t )}$ can be seen as the angle ranges of the paths from the circular region ${\bf{C}}^{(t-1 )}_0$ to predicted UE ${\bf{\hat x}}{'^{(t )}}$, as illustrated in Fig. \ref{e}(a). 
Thus, $\theta _{{\rm{scale,0}}}^{(t )},\phi _{{\rm{scale,0}}}^{(t )}$ can be calculated as
\begin{equation}
\theta _{{\rm{scale,0}}}^{\left(t \right)} =\phi _{{\rm{scale,0}}}^{\left(t \right)}= 2{\rm{arcsin}}\left({\frac{{R _0^{\left({t - 1} \right)}}}{{\left\| {\hat {\bf{x}}_0^{\left({t - 1} \right)} - \hat {\bf{x}}{'^{\left(t \right)}}} \right\|}_2 }} \right).\label{19}
\end{equation}

NLoS paths $(l\geq1)$ have different transmission characteristics compared with the LoS path.
NLoS paths first start from PA, then to the reflection point ${\bf{P}}$, and finally to the UE, as shown by the blue trajectory in Fig. \ref{e}(b). 
When VA is moving around on the circle of ${\bf{C}}^{(t-1 )}_l$, the point ${\bf{P}}$ formulates a closed convex ${\bf{P}}^{(t-1 )}_l$ in red. 
We set the slopes of the tangents to ${\bf{P}}^{(t-1 )}_l$ from ${\bf{\hat x}}{'^{(t )}}$ and ${\bf{\hat x}}_0^{({t - 1} )}$ as $\kappa_{l,{\rm{UE,max}}}^{(t)} > \kappa_{l,{\rm{UE,min}}}^{(t)}$ and $\kappa_{l,{\rm{PA,max}}}^{(t)} > \kappa_{l,{\rm{PA,min}}}^{(t)}$, respectively. 
Then, the definition of the prior angle information of the $l$-th path $\theta _{{\rm{VA}},l}^{(t )},$ $\phi _{{\rm{VA}},l}^{(t )},$ $\theta _{{\rm{scale}},l}^{(t )},$ $\phi _{{\rm{scale}},l}^{(t )}\in \left(0,\pi \right)$ is similar to those of the LoS path and can be calculated as
\begin{subequations}
\label{21}
\begin{align}
\theta _{{\rm VA},l}^{\left(t \right)} &= \frac{{{\mathop{\rm atan}\nolimits} \left({\kappa_{l,{\rm{UE,max}}}^{(t)}} \right) + {\rm atan} \left({\kappa_{l,{\rm{UE,min}}}^{(t)}} \right)}}{2} + 0.5\pi, \\
\phi _{{\rm VA},l}^{\left(t \right)} &= \frac{{{\rm atan} \left({\kappa_{l,{\rm{PA,max}}}^{(t)}} \right) + {\rm atan} \left({\kappa_{l,{\rm{PA,min}}}^{(t)}} \right)}}{2} + 0.5\pi, \\
\theta _{{\rm scale},l}^{\left(t \right)} &= {\rm atan} \left({\kappa_{l,{\rm{UE,max}}}^{(t)}} \right) - {\rm atan} \left({\kappa_{l,{\rm{UE,min}}}^{(t)}} \right),\\
\phi _{{\rm scale},l}^{\left(t \right)} &= {\rm atan} \left({\kappa_{l,{\rm{PA,max}}}^{(t)}} \right) - {\rm atan} \left({\kappa_{l,{\rm{PA,min}}}^{(t)}} \right).
\end{align}
\end{subequations}
In Fig. \ref{e}, the yellow regions illustrate the prior angle searching ranges centered on the prior angle values, which are believed to cover the true angles. 
\subsubsection{Feature-aided Tracking}
This sub-module utilizes prior angle information $\theta _{{\rm{VA}},l}^{(t )},$ $\phi _{{\rm{VA}},l}^{(t )},$ $\theta _{{\rm{scale}},l}^{(t )},$ $\phi _{{\rm{scale}},l}^{(t)}$,
$l = 0, ...,{\hat L}^{({t - 1} )} - 1$ to realize the overhead-reduced beam tracking. 
A total of ${\hat L}^{({t - 1} )}$ paths are tracked, and the tracking procedure of the $l$-th path is taken as an example. 
For brevity, the prior angle information of the $l$-th path is simplified as ${\phi_{\rm VA}}:=\phi _{{\rm{VA}},l}^{(t )} , {\phi _{{\rm{scale}}}}:= \phi _{{\rm{scale}},l}^{(t )},{\theta_{\rm VA}} := \theta _{{\rm{VA}},l}^{(t)}, {\theta _{{\rm{scale}}}}:=\theta _{{\rm{scale}},l}^{(t )}$ in this sub-section. 
The tracking procedure considers measuring beam pairs with different widths, similar to the hierarchical sweeping module. 
However, the center of the tracking codebook is focused on ${\phi_{\rm VA}},{\theta_{\rm VA}}$, and the initial tracking layer is decided by $\phi _{{\rm{scale}},l},\theta _{{\rm{scale}},l}$, as shown in Fig. \ref{d}(b).

{\color{black}{
The tracking codebook of the BS is taken for example, the $j$-th $(j=1,...,J)$ layer of the $l$-th path's codebook only has one subset ${ {{{\bf{F}}_{({j,{\phi_{\rm VA}} } )}}} }\in {\mathbb{C}^{{N_{{\rm BS}}} \times {2}}}$ consisting of two beamforming vectors. 
We define ${{\cal I}_{({j,{\phi_{\rm VA}},m} )}} = \{ { \frac{N}{{{2^j}}}({m - 2} ) + \lceil\frac{{\phi_{\rm VA}} }{\pi/N}\rceil, ...,\frac{N}{{{2^j}}}({m - 1} ) +\lceil\frac{{\phi_{\rm VA}} }{\pi/N}\rceil} \}$ centered on the discrete angle $\lceil\frac{{\phi_{\rm VA}} }{\pi/N}\rceil$, and ${{\bf{G}}_{({j,{\phi_{\rm VA}} } )}}\in {\mathbb{R}^{N \times {2}}}$ with the elements of the $m$-th column and the $u$-th row satisfying $u \in {{\cal I}_{({j,{\phi_{\rm VA}},m} )}}$ as 1, whereas others 0. Thus, ${\bf{F}}_{({j,{\phi_{\rm VA}} } )}$ can be similarly calculated according to (\ref{7}) and (\ref{8}). 
}}
{\color{black}{
Additionally, the tracking codebook $ {{{\bf{W}}_{({j,{\theta_{\rm VA}}} )}}}\in {\mathbb{C}^{{N_{{\rm UE}}} \times {2}}} $ at the UE side can be generated similarly.
}}

The tracking procedure can be simplified based on the tracking codebook. As the true AoD $\phi_{l}^{(t)}$ is believed to be in range $( \phi_{\rm VA}+ 0.5{\phi _{{\rm{scale}}}},\phi_{\rm VA}-0.5{\phi _{{\rm{scale}}}} )$, the initial tracking layer ${j^{\rm BS}}$ in BS can be calculated through the range width 
\begin{equation}\label{22}
{j^{\rm BS}} = \left\lceil{{{\log }_2}\frac{{{\phi _{\rm scale}}}}{{{\rm{\pi }}/N}}}\right\rceil,
\end{equation}
and the initial tracking layer ${j^{\rm UE}}$ in UE can be similarly calculated through $\theta _{{\rm{scale}}}$. 
Fig. \ref{d}(b) shows the tracking procedure. Finally, the channel parameter estimates of the $l$-th path can be given as (\ref{12}).
A total of ${\hat L}^{({t - 1} )}$ path in the descending order of $\hat g_l^{({t - 1} )}$ are tracked by repeating the tracking procedure through the successive interference cancellation, which is similar to the hierarchical sweeping module. 
Notably, the termination criterion in Section III-A is also considered at the beginning of each path's tracking procedure to check whether the path exists at time $t$. 
When the criterion is satisfied, the corresponding tracking procedure is omitted. 
Thus, the estimates of the feature-aided tracking module can be given as $\hat \theta _{l'}^{(t )},\hat \phi _{l'}^{(t )},\;\hat g_{l'}^{(t )},{l'} = 0,1, ...,{\hat L}^{(t )} - 1$ with ${{\hat L}^{({t} )}}={\hat L}^{({t - 1} )}$. 
{\color{black}A complete algorithm of the proposed feature-aided tracking module is presented in Algorithm 2.} 

The hierarchical sweeping and feature-aided tracking modules have different characteristics. 
The hierarchical sweeping module performs independently and covers the whole angle domain, which means this module requires heavy overhead. 
However, the hierarchical sweeping module can detect the birth or death of channel paths and adapt to changeable wireless environments.
Meanwhile, the feature-aided tracking module depends on the results from the sensing service and focuses on a specific angle domain, which means this module can reduce overhead consumption and noise effect. 
However, the feature-aided tracking module relies on prior angle information and cannot detect the birth of channel paths.
Therefore, the beam management should switch between the hierarchical sweeping and feature-aided tracking modules to make the most of their advantages.

\begin{algorithm}[t]
\caption{{\color{black}Feature-aided Tracking}}\label{A2}
\KwIn{Predicted UE location ${\bf{\hat x}}{'^{(t )}}$; Estimated radio map ${\bf{\hat x}}_l^{({t - 1} )},{\bf{\hat x}}_{l,p}^{(t-1 )}$ for $l = 0,1, ...,{\hat L}^{({t - 1} )} - 1$;}
Calculate prior angle information {\color{black}using} (17) and (21)\;
Set $l = 1$\;
\While{$l \leq {\hat L}^{(t-1)}$}{
Calculate partial channel estimation {\color{black}using} {(9)}\;
Calculate ${j^{\rm BS}}, {j^{\rm UE}}$ {\color{black}using} {(22)}\;
Set ${j}={\rm {min}}\left\{ {j^{\rm BS}}, {j^{\rm UE}} \right\}$\;
\For{$j \leq J$}
{
Run beam measurement {\color{black}using} {(10)}\;
Calculate residual signal {\color{black}using} {(11)}\;
Calculate the subset index {\color{black}using} {(12)} for the $\left(j+1\right)$-th layer\;
Set {$j= j+1$}\;
}
Run channel parameter estimation of the $l$-th path {\color{black}using} {(13)}\;
Set {$l= l+1$}\;
}
Set {${\hat L}^{(t )}= {\hat L}^{(t-1)}$}\;
\KwOut{Channel parameter estimation $\hat \theta _{l'}^{(t )},\hat \phi _{l'}^{(t )},\hat g_{l'}^{(t )}$ for $l'= 0,1, ...,{\hat L}^{(t )} - 1$;}
\end{algorithm}

\vspace{-2.5mm}
\subsection{Switching Module}
The switching module is shown in the gray region of Fig. \ref{c}, consisting of two sub-modules. 
The first sub-module is set at the beginning of the beam management, deciding whether the hierarchical sweeping or feature-aided tracking module is performed.
The second one is set after the feature-aided tracking module, which is used to judge the effectiveness of the feature-aided tracking module and decide whether the hierarchical sweeping module is required additionally. 

These two sub-modules share the same design. First, the channel power is calculated through the estimated channel parameters. Then, the calculated channel power is compared with the historical one for CSI abrupt change detection. 
The channel power calculation and the CSI abrupt change detection are introduced as follows. 

\subsubsection{Channel Power Calculation} 
For CSI abrupt change detection, the channel power calculated through the estimated channel parameters in (\ref{12}) from the hierarchical sweeping or feature-aided tracking module should be introduced first.
Based on the estimated parameters, the channel estimation ${{\bf{\hat H}}}^{\left(t \right)} \in {\mathbb{C}^{{N_{\rm UE}} \times {{N_{\rm BS}}}}} $ at time $t$ can be calculated as
\begin{equation}
\label{24}
{{{{\bf{\hat H}}}^{\left(t \right)}}}=\mathop \sum \limits_{l = 0}^{{{\hat L}^{\left({t} \right)}}-1} \hat g_l^{\left(t \right)}{{\bf{a}}_{{\rm{UE}}}}\left({\hat \theta _l^{\left(t \right)}} \right){\bf{a}}_{{\rm{BS}}}^{\rm H}\left({\hat \phi _l^{\left(t \right)}} \right).
\end{equation}
However, the channel power calculation based on ${{\bf{\hat H}}}^{(t )}$ is mainly influenced by the LoS path. 
The birth or death of NLoS paths may be swamped because of the LoS path gain change in mmWave systems. Therefore, the LoS and NLoS paths should be considered separately.

The LoS path CSI component ${\bf{\hat H}}_{{\rm{LoS}}}^{(t )} \in {\mathbb{C}^{{N_{\rm UE}} \times {{N_{\rm BS}}}}}$ can 
be observed through its significantly high path gain, which is given by
\begin{subequations}
\label{25}
\begin{align}
\label{25a}
{\bf{\hat H}}_{{\rm{LoS}}}^{\left(t \right)} &={e_{\rm LoS}^{\left(t \right)}}{ \hat g_0^{\left(t \right)}{{\bf{a}}_{{\rm{UE}}}}\left({\hat \theta _0^{\left(t \right)}} \right){\bf{a}}_{{\rm{BS}}}^{\rm{H}}\left({\hat \phi _0^{\left(t \right)}} \right)},
\\
\label{25b}
e_{{\rm{LoS}}}^{\left(t \right)} &= \left\{ {\begin{array}{*{20}{l}}
{1,\;{\rm{if}}\mathop {\max }\limits_l \hat g_l^{\left(t \right)} > {g_{\rm LoS}}},\\
{0,\;{\rm{otherwise}}},
\end{array}} \right.
\end{align}
\end{subequations}
\noindent where $e_{{\rm{LoS}}}^{(t )}$ is a $0{\raisebox{0mm}{-}}1$ variable describing the existence of the LoS path, and ${g_{\rm LoS}}$ is the minimum LoS path gain threshold. 

Based on (\ref{24})-(\ref{25}), the channel power ${E^{(t )}}$ in \cite{a14} can be calculated by setting the LoS path power as $e_{{\rm{LoS}}}^{(t )}\Delta _E^{\rm min}$ with constant $\Delta _E^{\rm min}$. 
Then, the corresponding power difference $\Delta _E^{({t,t - 1} )}$ is set to measure the difference between ${E^{(t )}}$ and ${E^{(t-1 )}}$.
They can be calculated as
\begin{subequations}\label{26}
\begin{align}
{E^{\left(t \right)}} &= {\left\| {{{ {{{{\bf{\hat H}}}^{\left(t \right)}} - {\bf{\hat H}}_{{\rm{LoS}}}^{\left(t \right)}}}}} \right\|^2_F} + e_{{\rm{LoS}}}^{\left(t \right)}\Delta _E^{\rm min},\\
\Delta _E^{\left({t,t - 1} \right)} &=\frac{ \left| {{E^{\left(t \right)}} - {E^{\left({t - 1} \right)}}} \right|}{{E^{\left({t - 1} \right)}}}.
\end{align}
\end{subequations}
\noindent The path gain change of LoS path is neglected when its existence is detected through $e_{{\rm{LoS}}}^{(t )}$, which allows $\Delta _E^{({t,t - 1} )}$ to detect the birth or death of NLoS paths clearly when LoS path exists.

\subsubsection{CSI Abrupt Change Detection}
After the channel power calculation in (\ref{26}), the sub-modules detect the CSI abrupt change. The logical variable ${D^{(t)}}\in \{ {\rm{True}},{\rm{False} }\}$ is set to represent whether CSI abrupt change happens or not.
For detection before the beam management sub-module, ${D^{(t-1)}}$ is decided through ${E^{(t-1 )}}$ and $\Delta _E^{({t-1,t - 2} )}$, which compare the channel power at time $t-1$ with that at time $t-2$.
For detection after the tracking sub-module, ${D^{(t)}}$ is decided by comparing the channel power calculated at time $t$ with that at time $t-1$. 
The channel power at time $t$ is calculated utilizing the estimated parameters of the feature-aided tracking module. 

The calculation of ${D^{(t)}}$ is taken for example, where ${E^{(t)}}$ is set to measure the channel power, and $\Delta _E^{({t,t - 1} )}$ is set to detect the birth or death of NLoS paths. 
When channel power is too small or channel power difference is too large, ${D^{(t)}}$ is set as true, and not otherwise. Thus, the calculation can be given by
\begin{equation}\label{28}
{D^{(t)}} = \left\{ {\begin{array}{*{20}{l}}
{{\rm{True}},\;
{E^{\left(t \right)}} < {E_{{\rm{min}}}}}
\;{\rm{ or }}\;
\Delta _E^{\left({t ,t - 1} \right)} > \Delta _E^{\rm min}
,\\
{{\rm{False,\;otherwise}}},
\end{array}} \right.
\end{equation}
\noindent where $\Delta _E^{\rm min}$ is a constant threshold corresponding to the minimum power threshold of the CSI abrupt change, and ${E_{{\rm{min}}}}$ is the minimum channel power threshold satisfying the UE's communication demands. 
When $\Delta _E^{({t ,t - 1} )}$ is beyond the threshold $\Delta _E^{\rm min}$, or ${E^{(t )}}$ cannot reach the threshold ${E_{{\rm{min}}}}$, and the CSI abrupt change is detected (${D^{(t )}}$ is true). 

As shown in Fig. \ref{c}, if an abrupt change in CSI is detected (${D^{(t-1 )}}$ is true) in the detection before beam management sub-module, the hierarchical sweeping module is chosen to perform sweeping across the full-angle domain. 
Otherwise, the feature-aided tracking module is selected for tracking with reduced overhead. {\color{black}Meanwhile, if a CSI abrupt change is detected (${D^{(t )}}$ is true) in the detection after tracking sub-module, the hierarchical sweeping module is required to compensate for the feature-aided tracking module. Otherwise, the hierarchical sweeping module is not necessary for compensation.} 
The switching module completes the joint beam management and SLAM design, as depicted in Fig. \ref{c}. Algorithm 3 provides a summary of the proposed beam management scheme, which encompasses the hierarchical sweeping module, the feature-aided tracking module, and the switching module.

\begin{algorithm}[t]
\caption{{\color{black}Beam Management with Switching Module}}\label{A3}
\KwIn{ Switching logical symbol $D^{(t-1)}$;}
\eIf{$D^{(t-1)}={\rm{True}}$}
{Run Algorithm 1\;
}
{
Run Algorithm 2\;
Calculate $D^{(t)}$ {\color{black}using} {(26)} based on the results from Algorithm 2\;
\If{$D^{(t)}={\rm{True}}$}
{{\color{black}Rerun Algorithm 1 to compensate for Algorithm 2}\;
}
}
Update $D^{(t)}$ {\color{black}using} {(26)} based on the results from Algorithm 1 or 2\;
\end{algorithm}

% Performance Metric
\vspace{-3mm}
\section{Performance Metric}
The proposed joint design can obtain estimates including channel parameters $\hat \theta _l^{(t )},\hat \phi _l^{(t )},\hat g_l^{(t )},l = 0,1, ...,{\hat L}^{(t )} - 1$, channel ${{{{\bf{\hat H}}}^{(t )}}}$, PA/VA locations ${\bf{\hat x}}_l^{(t )},l = 0,1, ...,{\hat L}^{(t )} - 1$, and UE location ${\bf{\hat x}}^{(t )}$. 
Therefore, this section introduces different performance metrics to evaluate the proposed joint design from the aspects of sensing and communication. 
The results at time $t$ are considered, and the superscript $(t)$ is omitted for simplicity in this section.
\vspace{-3mm}
\subsection{Sensing Performance Metrics}
The performance metrics for sensing can be divided into two types.
On the one hand, the sensing ability of beam management is evaluated by the angle estimation error ${{\rm{E}}_{{\rm{angle}}}}$.
{\color{black}Considering that the estimated path number ${\hat L}$ may mismatch the true value ${L}$, the metric ${{\rm{E}}_{{\rm{angle}}}}$ should consider both the angle estimation accuracy and the path number estimation accuracy. }
Therefore, ${{\rm{E}}_{{\rm{angle}}}}$ can be designed as the multi-object miss-distance metric describing the difference between finite nonempty subsets $\{ [ {{\hat \theta }_1} , {{\hat \phi }_1}]^{\rm T} ,..., [ {{\hat \theta }_{\hat L}}, {{\hat \phi }_{\hat L}}]^{\rm T} \}$ and $\{ [ {{ \theta }_1} , {{ \phi }_1}]^{\rm T} ,..., [ {{ \theta }_{ L}}, {{ \phi }_{ L}}]^{\rm T} \}$. We define the miss-distance $d_{\rm angle}\left( l,l' \right)$ between $[ {{ \theta }_l} , {{ \phi }_l}]^{\rm T}$ and $[ {{\hat \theta }_{l'}}, {{\hat \phi }_{l'}}]^{\rm T}$ as 
\begin{equation}
d_{\rm angle}\left( l,l' \right)=\left| {\theta _{l}}-{{\hat \theta }_{l'}} \right| + \left| {\phi _{l}}-{{\hat \phi }_{l'}} \right| .
\end{equation}
\noindent The metric ${{\rm{E}}_{{\rm{angle}}}}$ can be calculated according to \cite{a16} 
\begin{equation}
\label{29}
{{\rm{E}}_{{\rm{angle}}}} = \frac{1}{{2{L_{{\rm{max}}}}}}\left({\mathop {\min }\limits_{\varepsilon \in {{\rm{\Pi }}_{{L_{{\rm{eff}}}}}}} \mathop \sum \limits_{l = 0}^{{L_{{\rm{eff}}}} - 1}
d_{\rm angle}\left( {\varepsilon \left(l \right)},l \right)+ 2{c_{{\rm{angle}}}}{L_{\rm{\Delta }}}} \right),
\end{equation}
\noindent where ${L_{\max }} = {\rm{max}}\{ {L,{{\hat L}}} \}$, ${L_{{\rm{eff}}}} = {\rm{min}}\{ {L,{{\hat L}}} \}$, and ${L_\Delta } = | {L - {{\hat L}}} |$. The set ${{\rm{\Pi }}_{{L_{{\rm{eff}}}}}}$ represents the permutations on $\{0,1,...,{{\hat L}}-1\}$.
The penalty term $2{c_{{\rm{angle}}}}{L_\Delta }$ with constant ${c_{{\rm{angle}}}}$ is considered to measure the path number estimation error.

On the other hand, the sensing service provided by SLAM is evaluated by the estimation error of UE location ${{\rm{E}}_{{\rm{UE}}}}$ represented as
\begin{equation}
\label{30}
{{\rm{E}}_{{\rm{UE}}}} = {\left\| {{\bf{\hat x}} - {\bf{x}}} \right\|}_2,
\end{equation}
and the radio map construction error is represented by optimal sub-pattern assignment (OSPA) in \cite{a6} as 
\begin{equation}
\label{31}
{\rm{OSPA}} = \frac{1}{{{L_{{\rm{max}}}}}}\left({\mathop {\min }\limits_{\varepsilon \in {{\rm{\Pi }}_{{{L_{{\rm{eff}}}}}}}} \mathop \sum \limits_{l = 0}^{{L_{{\rm{eff}}}} - 1} 
\left\|{{\hat {\bf {x}}}_l} - {\bf{x}}_{\varepsilon \left(l \right)}\right\|_2 
+ {c_{{\rm{map}}}}{L_{\rm{\Delta }}}} \right),
\end{equation}
\noindent where ${c_{{\rm{map}}}}$ is the penalty constant similar to ${c_{{\rm{angle}}}}$ in (\ref{29}).
\vspace{-3mm}
\subsection{Communication Performance Metrics}
Three performance metrics for communication are considered in this study.
First, the channel estimation is evaluated by normalized mean squared error (NMSE) represented as 
\begin{equation}
\label{32}
{\rm{NMSE}} = {{ {\left\| {\bf{H}} - {\bf{\hat H}} \right\|}_F^2 }}\bigg/{{ {\left\| {\bf{H}} \right\|}_F^2 }}.
\end{equation}

Then, the beam management overhead can be evaluated utilizing the total layers used by hierarchical sweeping or feature-aided tracking modules. 
For the hierarchical sweeping module, the layer number for each path is fixed as $J$. 
For the feature-aided tracking module, the layer number of the $l$-th path is given by $(J-{j_l}+1)$ with ${j_l} = \min \{ j_l^{{\rm{BS}}},j_l^{{\rm{UE}}}\} $ in (\ref{24}). 
For each layer, UE and BS consider two different beams, and a total of four different beam pairs are measured. Thus, the beam management overhead ${T}_{\rm BM}$ can be calculated as
\begin{equation}
\label{33}
{{{T}}_{{\rm{BM}}}} = \left\{ {\begin{array}{*{20}{l}}
4{J{{\hat L}}{ T_{\rm b}},\;\;\;\;\;\;\;\;\;\;\;\;\;\;\;\;\;\;\;\;\;\;\;\;\;{\mbox{for hierarchical sweeping}}},\\
{\mathop \sum \nolimits_{l = 0}^{{{\hat L}}-1} 4 \left({J - {j_l}}+1 \right){ T_{\rm b}},\,\,{\mbox{for feature-aided tracking}}},
\end{array}} \right. 
\end{equation}
\noindent where ${ T_{\rm b}}$ represents the time for measuring a single beam pair.

Finally, the corresponding SE is calculated to evaluate the communication ability synthetically. For calculation, the joint precoding design based on ${{\bf{\hat H}}}$ is discussed as follows. 
SE with perfect CSI $\bf H$ can be calculated as \cite{a1601}
\begin{equation}
\label{34}
{\rm{S}}{{\rm{E}}_{\rm{P}}} = {\log _2}\left({ {1} + \frac{{\left|{{\bf{w}}^{\rm H}{\bf{H}}{{\bf{f}}}} \right|}^2 }{{\sigma ^2}} } \right).
\end{equation}
\noindent The precoding problem can be decoupled into two steps in \cite{a17}: Design of the precoder ${{\bf{f}}}$ in BS, followed by the combiner design ${{\bf{w}}}$ in UE. For BS, the precoder design ${{{\bf{ f}}}_{{\rm{opt}}}}\in \mathbb{C}^{{N_{{\rm{BS}}}} \times {1}}$ can be described as
\begin{equation}
{{{\bf{ f}}}_{{\rm{opt}}}} = \mathop {{\rm{arg\;max}}}\limits_{{{\bf{f}}}} {\cal I}\left({{{\bf{f}}}} \right),
\end{equation}
\noindent where ${\cal I}({{{\bf{f}}}} ) = {\log _2}({| {{\bf{I}} + \frac{{\rm{1 }}}{{{{\rm{\sigma }}^2}}}{\bf{H}}{{\bf{f}}}{\bf{f}}^{\rm H}{{\bf{H}}^{\rm H}}} |} )$ represents the mutual information of Gaussian signals on $\bf{H}$. 
The channel’s singular value decomposition (SVD) is considered as ${\bf{H = U\Sigma }}{{\bf{V}}^{\rm H}}$ with the unitary matrix ${\bf{U}} \in \mathbb{C}^{{N_{{\rm{UE}}}} \times {\rm{rank}}({\bf{H}} )}$, 
the diagonal matrix ${\bf{\Sigma }} \in \mathbb{C}^{{\rm{rank}}({\bf{H}} ) \times {\rm{rank}}({\bf{H}} )}$, 
and the unitary matrix ${\bf{V}} \in \mathbb{C}^{{N_{{\rm{BS}}}} \times {\rm{rank}}({\bf{H}} )}$. 
Furthermore, we define the first column of $\bf{V}$ as the vector ${{\bf{v}}_1}\in \mathbb{C}^{{N_{{\rm{BS}}}} \times {1}}$. 
Thus, the optimal unconstrained unitary precoder for $\bf{H}$ is simply given by ${{{\bf{ f}}}_{{\rm{opt}}}} = {{{\bf{ v}}}_1}$.
As for UE, the combiner design ${{\bf{w}}_{{\rm{MMSE}}}}\in \mathbb{C}^{{N_{{\rm{UE}}}} \times {1}}$ can be transformed into a minimum mean square error (MMSE) problem as
\begin{equation}
{{\bf{w}}_{{\rm{MMSE}}}} = \mathop {{\rm{arg\;max}}}\limits_{{{\bf{w}}}} 
\mathbb{E}
\left[{
{\left| { {{{s}} - {\bf{w}}^{\rm H}\left({{\bf{H}}{{{\bf{ f}}}_{{\rm{opt}}}}+{\bf n}}\right)} } \right|}^2
}\right]
,
\end{equation}
which can be solved as
\begin{equation}
\label{37}
{\bf{w}}_{{\rm{MMSE}}}^{\rm H} = 
{\left({\left\| \bf{H}{{\bf{f}}_{{\rm{opt}}}}\right\|_2^2 + {{{\rm{\sigma }}^2}}} \right)^{ - 1}}{\bf{f}}_{{\rm{opt}}}^{\rm H}{{\bf{H}}^{\rm H}}.
\end{equation}
Based on the estimated CSI ${{\bf{\hat H}}}$, the SVD precoder ${{{{\bf{\hat f}}}_{{\rm{opt}}}}}={{{\bf{\hat v}}}_1}$ can be constructed as the first column of ${{\bf{\hat V}}}$ from $\hat {\bf{H}} = {\bf{\hat U\hat \Sigma }}{{{\bf{\hat V}}}^{\rm H}}$, and the MMSE combiner ${\bf{\hat w}}_{{\rm{MMSE}}}$ can be similarly calculated as (\ref{37}). 
After the precoding design, we set the corresponding effective channel as $h_{\rm eff}={\bf{\hat w}}_{{\rm{MMSE}}}^{\rm H} {\bf H} {\bf{\hat f}}_{\rm{opt}}$. We consider transmit $|{ s}|=1$ through beam pair ${\bf{\hat w}}_{{\rm{MMSE}}}$ and ${\bf{\hat f}}_{\rm{opt}}$, the corresponding received signal ${r}$ can be given by 
\begin{equation}
{r}=h_{\rm eff} s+{\bf{\hat w}}_{{\rm{MMSE}}}^{\rm H}{\bf n}.
\end{equation}
By repeating such transmission for adequate times, the effective channel estimation ${\hat h}_{\rm eff}$ can be calculated by averaging ${r}$.
Thus, ${\rm{S}}{{\rm{E}}_{{\rm{IP}}}}$ under the imperfect CSI $\hat {\bf{H}}$ can be calculated as
\begin{equation}
\label{40}
\begin{aligned}
{\rm{S}}{{\rm{E}}_{{\rm{IP}}}} = \frac{{{{T}} - {{{T}}_{{\rm{BM}}}}}}{{{T}}}\times{\log _2}\left( {{{1}} + \frac{\left|{\hat h}_{\rm eff}\right|^2 }{{{\rm{\sigma }}^2}+\left|{h}_{\rm eff}- {\hat h}_{\rm eff}\right|^2} }\right),
\end{aligned}
\end{equation}

In this section, the angle estimation error ${{\rm{E}}_{{\rm{angle}}}}$, UE localization error ${{\rm{E}}_{{\rm{UE}}}}$, and radio map construction error OSPA are introduced to evaluate the sensing performance. 
Meanwhile, the channel estimation error NMSE, the overhead consumption ${{{T}}_{{\rm{BM}}}}$, and ${\rm{S}}{{\rm{E}}_{{\rm{IP}}}}$ based on the imperfect CSI are introduced to evaluate the communication performance.
\begin{table*}[]
\caption{\color{black}Parameters for Joint Design Simulations}\vspace{-8mm}
\label{table2}
\center
\begin{tabular}{|l|l|l|l|}
\hline
SNR & SNR\;$=15\;{\rm {dB}}$ & Codebook Resolution & $N=512$ \\ \hline
IMU Noise & $\sigma _{{\rm{IMU}}}=0.02\;{\rm m/s^2}$ & Angle Searching Range Constant & ${{\rm C}_{\rm angle}}=3.27$ \\ \hline
\multirow{2}{*}{Antenna Number} & ${N_{{\rm{BS}}}} = 32$ & \multirow{2}{*}{Time Duration} & ${{T}} = 1\;{{\rm s}}$ \\ \cline{2-2} \cline{4-4} 
& ${N_{{\rm{UE}}}} = 16$ & & ${{{T}}_{\rm{b}}} = {1}/{{16}}\;{\rm{ms}}$ \\ \hline
\end{tabular}
\vspace{-10mm}
\end{table*}

% Numerical Results
\vspace{-3.5mm}
\section{Numerical Results}
In this section, we first perform the proposed joint design in different scenarios for evaluating the beam management and sensing service in changeable and stable wireless environments. 
We then discuss numerical results evaluating the performance of the hierarchical sweeping, feature-aided tracking, and switching modules. 
%Then, numerical results evaluating the hierarchical sweeping, feature-aided tracking and switching modules are discussed. 
\vspace{-5mm}
\subsection{Performance of Beam Management and SLAM Joint Design}
\begin{figure}
\centering
\subfloat[]{
\hspace{-3.8mm}\includegraphics[width=0.7\textwidth,trim=5 5 5 5,clip]{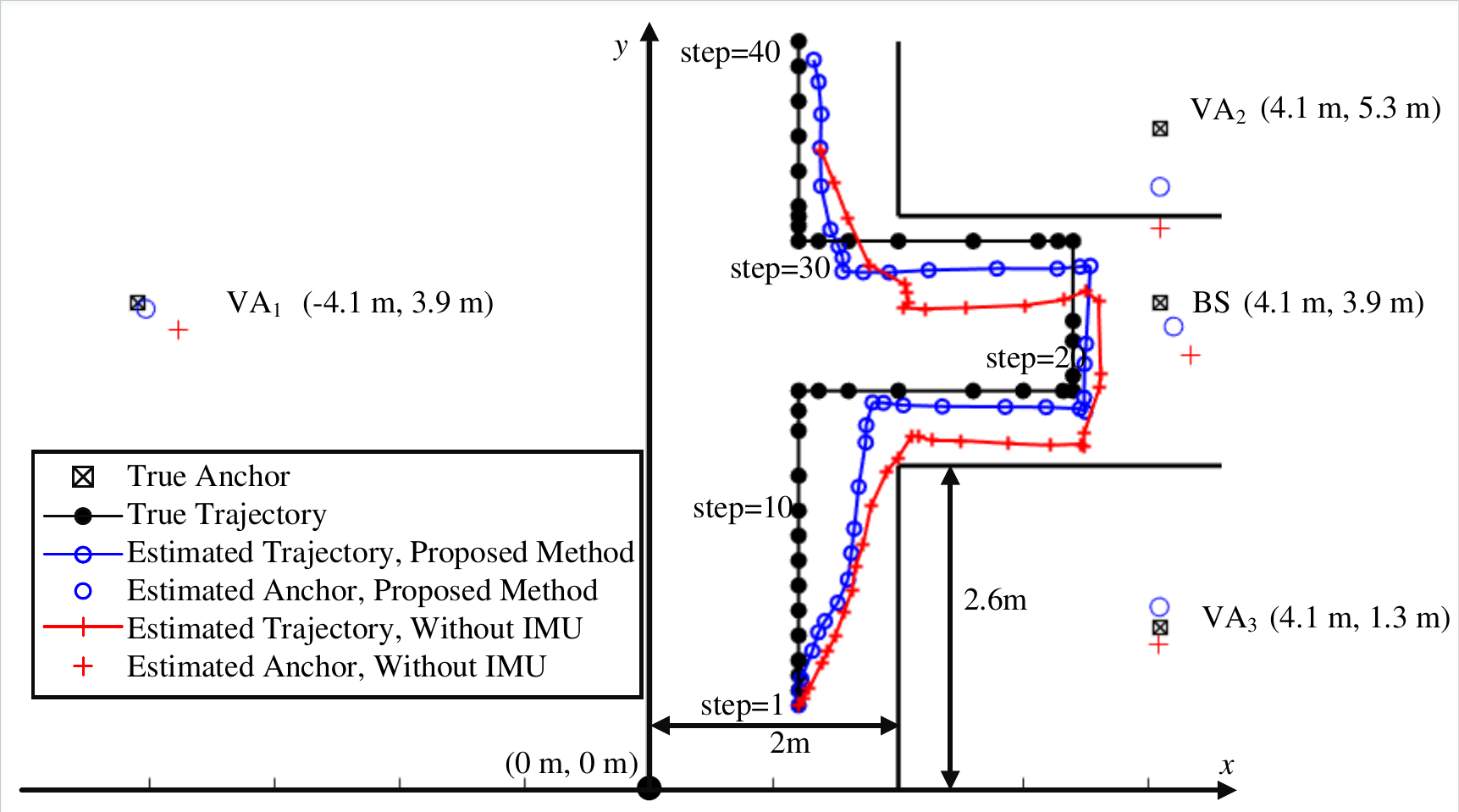}\hspace{-0mm}\label{g1}
}
\quad
\subfloat[]{
\hspace{-7.6mm}\includegraphics[width=0.7\textwidth,trim=5 5 5 5,clip]{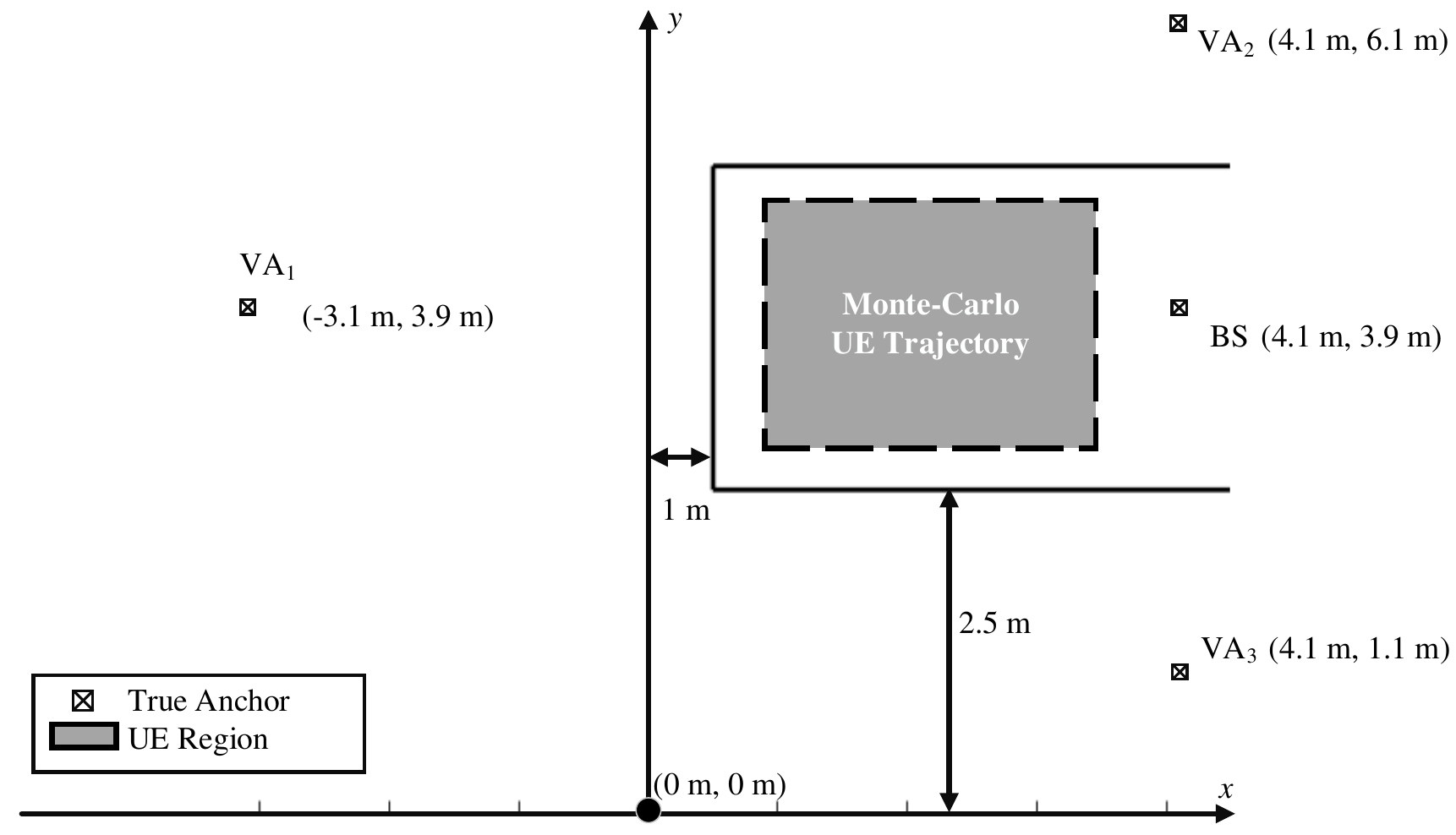}\hspace{-3mm}
}
\caption{Floor plan for the joint design simulations with different wireless environments: (a) Scenario 1 and (b) Scenario 2.}
\label{g} 
\vspace{-10mm}
\end{figure}

\vspace{-1.5mm}
This section analyzes the experiments for scenarios where UE moves in specific physical indoor environments. 
The experimental scenes are established at the 2D T-shaped intersection and a square hall with a single UE and BS, as shown in Fig. \ref{g}. 
TABLE \ref{table2} presents the parameters of this section. 
The proposed joint design performs with time interval $T=1{\;\rm s}$, and a single beam pair is measured for ${{{T}}_{\rm{b}}} = {1}/{{16}}\;{\rm{ms}}$.
In addition, the path angles are generated by (\ref{4}) based on the geometry environment, and the path gains are generated through the ray tracing path loss model in \cite{a18}. 
Three experiments are considered for evaluating the benefits of IMU and switching modules:
``Proposed Method'' represents the proposed joint design;
``Hierarchical Method'' considers the absence of the switching module and only performs the hierarchical sweeping module for beam management; 
``Without IMU'' considers the absence of IMU, where the predicted UE location at time $t$ is replaced by ${{\bf{\hat x}}^{({t - 1} )}}$ in (\ref{15}) and SLAM is performed without IMU. 
The performance metrics are set according to Section IV: ${{\rm{E}}_{{\rm{angle}}}}$ in (\ref{29}) with $c_{\rm angle}=0.1\;{\rm rad}$, ${\rm{S}}{{\rm{E}}_{{\rm{IP}}}}$ in (\ref{40}), and ${{{T}}_{{\rm{BM}}}}$ in (\ref{33}) are set for evaluating beam management; ${{\rm{E}}_{{\rm{UE}}}}$ in (\ref{30}) and OSPA in (\ref{31}) with $c_{\rm map}=3.5\;{\rm m}$ are set for evaluating sensing services. 
Two different scenarios in Fig. \ref{g} considering different wireless environments are evaluated. They are discussed as follows.

\subsubsection{Changeable Wireless Environment}
{\color{black}Scenario 1 considers the changeable wireless environment, as shown in Fig. \ref{g}(a). 
In this scenario, the UE first enters the BS servicing area from the south of the intersection, which causes an increase in the channel path number. 
The UE then moves to the central area, where the channel path number remains unchanged. Finally, the UE leaves the intersection, and the channel path number decreases.} 
Fig. \ref{g}(a) also show the results of ``Proposed Method’’ and ``Without IMU’’ through Scenario 1. 
The UE trajectory is fixed as the black curve. 
The blue curves and marks represent the results of ``Proposed Method’’, whereas the red ones represent the results of ``Without IMU’’. 
For Scenario 1, the blue trajectory is closer to the black trajectory compared with the red one. ``Proposed Method’’ has an average ${{\rm{E}}_{{\rm{UE}}}}$ reduction of $0.35\;{\rm m}$ compared with ``Without IMU’’. 
The blue circles are closer to the true anchors compared with the red crosses for VA${_1}$, VA${_2}$, and BS, with OSPA reduction of $0.3\;{\rm m}$. 
%Thus, the introduction of IMU can improve UE localization and radio map construction accuracy in the changeable wireless environment. 
\begin{figure}
\centering
\subfloat[]{
\hspace{-4.5mm}\includegraphics[scale=0.46]{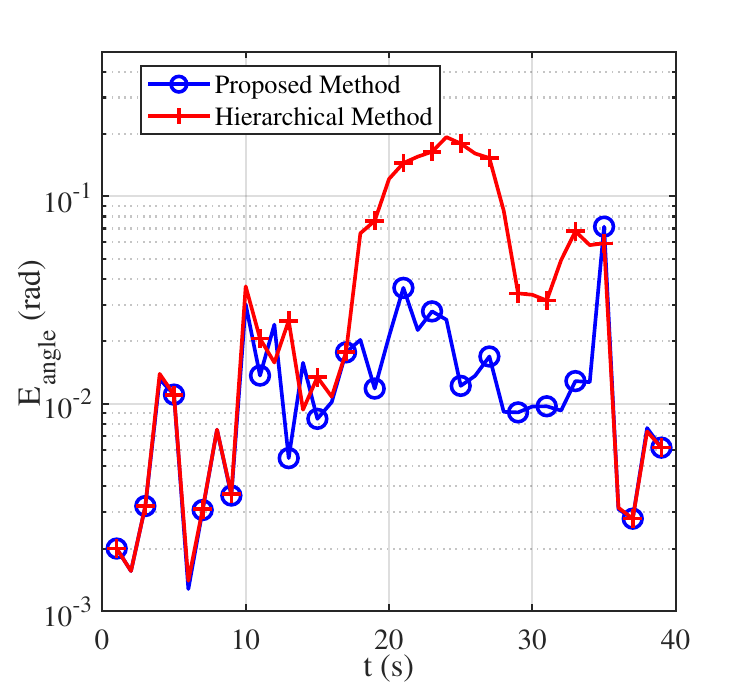}\hspace{-5mm}
}
\quad
\subfloat[]{
\hspace{-6mm}\includegraphics[scale=0.46]{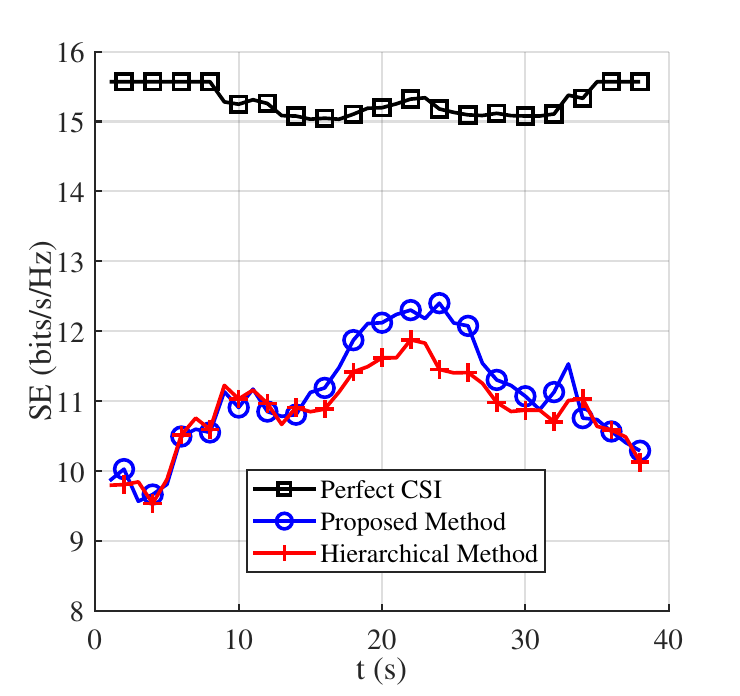}\hspace{-7mm}
}
\quad
\subfloat[]{
\hspace{-4.5mm}\includegraphics[scale=0.46]{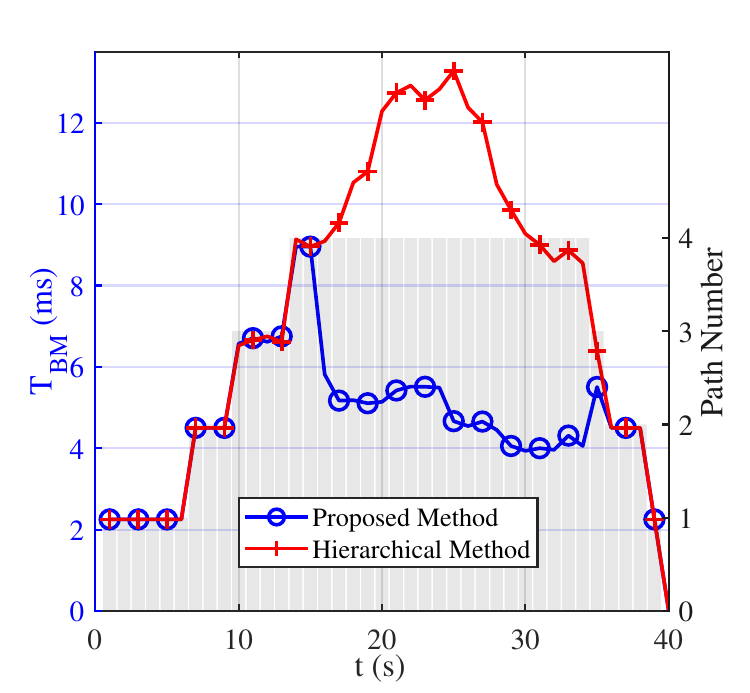}\hspace{-5mm}
}
\quad
\subfloat[]{
\hspace{-0mm}\includegraphics[scale=0.46]{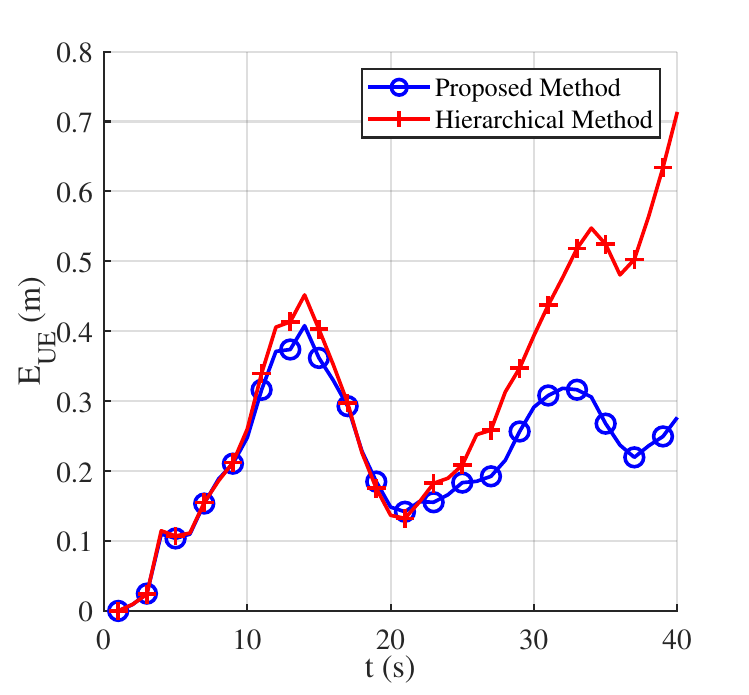}\hspace{-0mm}
}
\quad
\subfloat[]{
\hspace{-0mm}\includegraphics[scale=0.46]{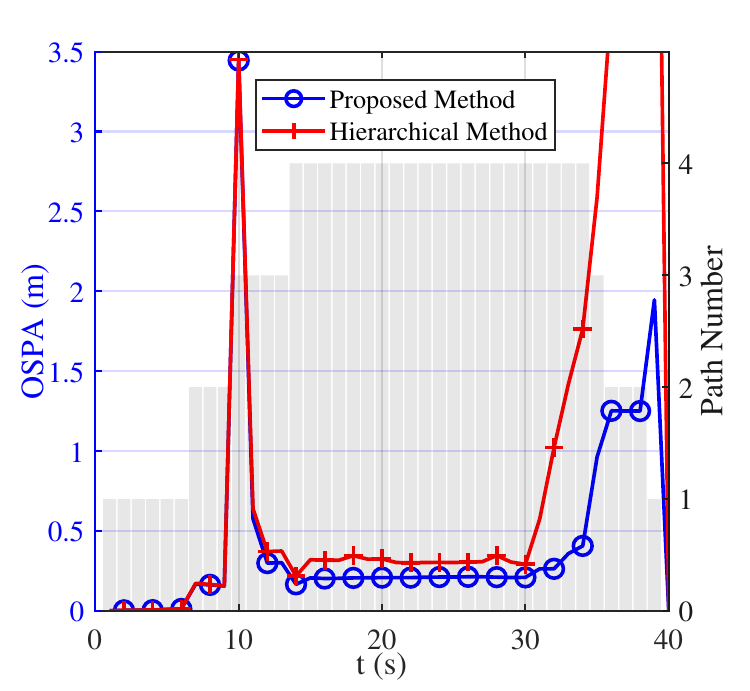}\hspace{-0mm}
}
\caption{
(a) Angle estimation performance, (b) SE performance, and (c) overhead consumption of the beam management; (d) UE localization accuracy and (e) radio map construction accuracy for ``Proposed Method’’ and ``Hierarchical Method’’ in Scenario 1. 
}
\label{h}
\vspace{-10mm}
\end{figure}

\begin{figure}
\centering
\subfloat[]{
\hspace{-4.5mm}\includegraphics[scale=0.46]{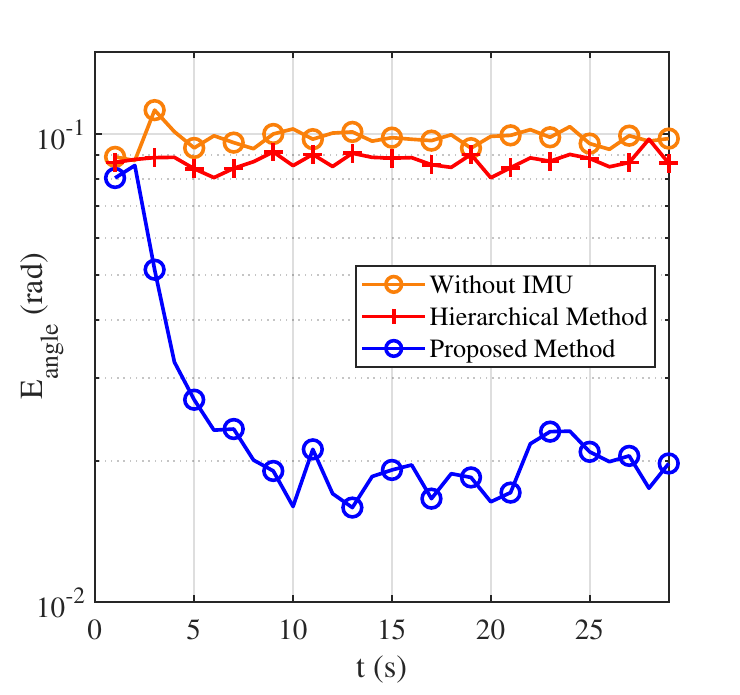}\hspace{-5mm}
}
\quad
\subfloat[]{
\hspace{-6mm}\includegraphics[scale=0.46]{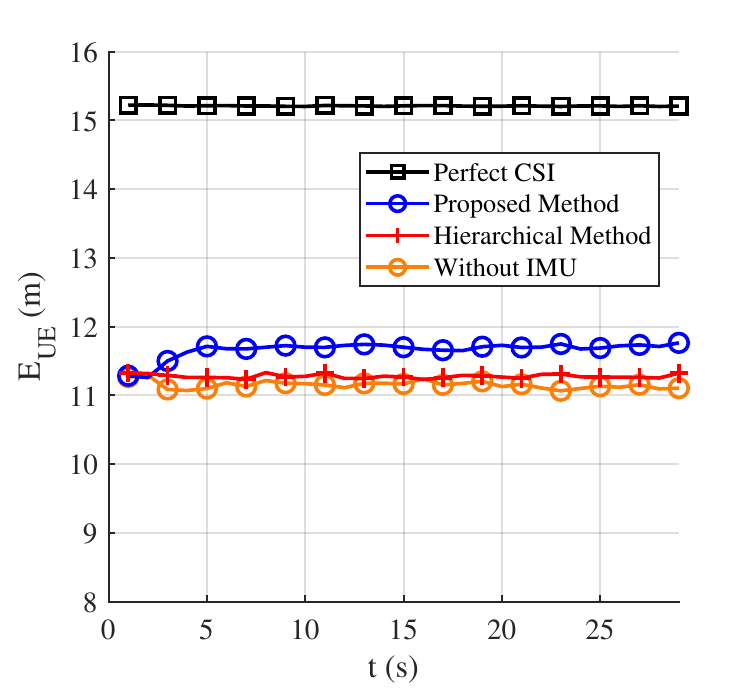}\hspace{-7mm}
}
\quad
\subfloat[]{
\hspace{-4.5mm}\includegraphics[scale=0.46]{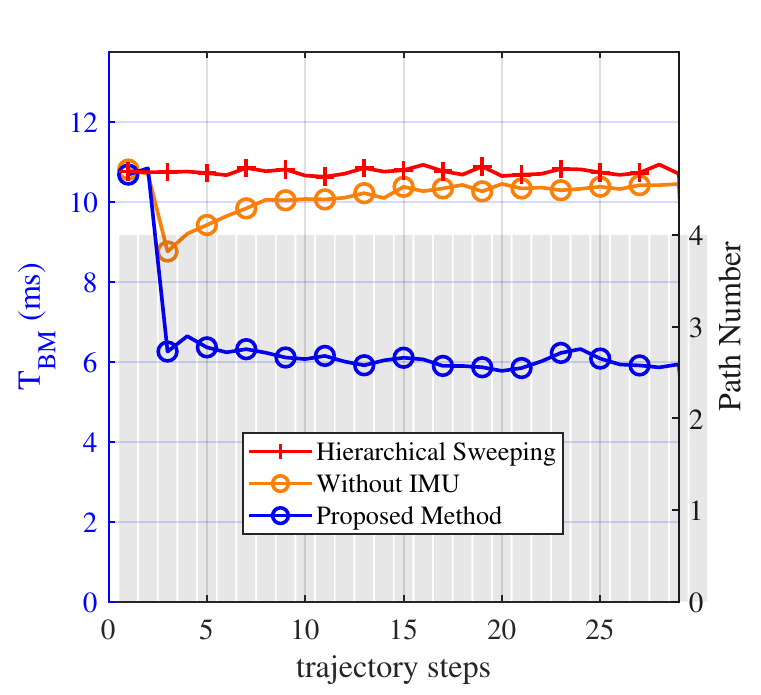}\hspace{-5mm}
}
\quad
\subfloat[]{
\hspace{-0mm}\includegraphics[scale=0.46]{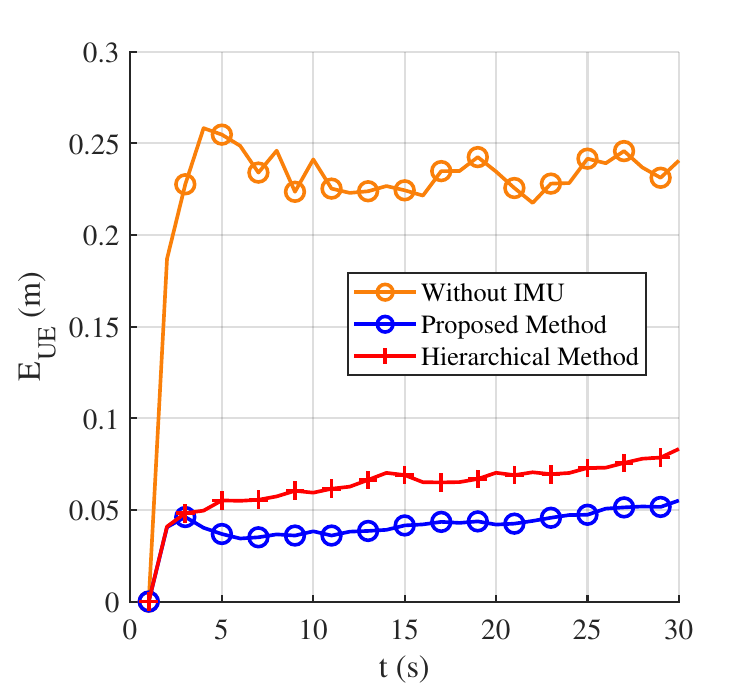}\hspace{-0mm}
}
\quad
\subfloat[]{
\hspace{-0mm}\includegraphics[scale=0.46]{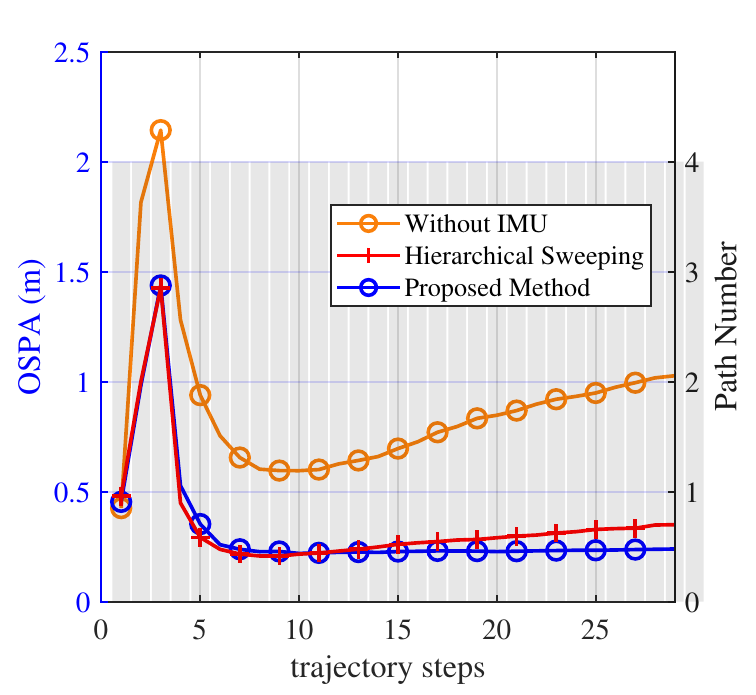}\hspace{-0mm}
}
\caption{
(a) Angle estimation performance, (b) SE performance, and (c) overhead consumption of beam management; (d) UE localization accuracy and (e) radio map construction accuracy for ``Without IMU'', ``Proposed Method’’, and ``Hierarchical Method’’ in Scenario 2.
}
\label{i}
\vspace{-10mm}
\end{figure}

Fig. \ref{h} shows the results of ``Proposed Method’’ and ``Hierarchical Method’’ with 5000 Monte-Carlo simulations in Scenario 1, where panels (a)-(c) show ${{\rm{E}}_{{\rm{angle}}}}$, ${\rm{S}}{{\rm{E}}_{{\rm{IP}}}}$, and ${{{T}}_{{\rm{BM}}}}$ trend diagrams evaluating the beam management, and panels (d)-(e) show ${{\rm{E}}_{{\rm{UE}}}}$ and OSPA trend diagrams evaluating the sensing services, respectively.
The red curves represent ``Hierarchical Method’’, whereas the blue ones represent ``Proposed Method’’. In addition, the gray bar describes the change of path number $L^{(t)}$. 
The analysis of the figure is as follows. 

First, Fig. \ref{h}(a) shows the trend diagram of the angle estimation error. 
The similar performance can be observed at $t \in {(1,10) }\cup {(35,40)}{\;\rm s}$, where path number $L^{(t)}<3$. 
The reason is that the experiments all perform the hierarchical sweeping module because of low channel power.
Meanwhile, ``Proposed Method’’ outperforms ``Hierarchical Method’’ at $t \in {(10,35) }{\;\rm s}$. 
This is when the path number $L^{(t)}=4$, and the feature-aided tracking module is activated by the switching module in ``Proposed Method’’, leading to the better performance of ``Proposed Method’’ compared with ``Hierarchical Method’’.
{\color{black}Fig. \ref{h}(b) and (c) illustrate that when the feature-aided tracking module is activated ($t \in {(10,35) }{\;\rm s}$), ``Proposed Method’’ can improve SE by $4\%$ while reducing overhead consumption by $56\%$, compared to ``Hierarchical Method’’.} 
The reason is that the feature-aided tracking module with effective prior angle information outperforms the hierarchical sweeping module with less overhead. 
The path number estimation error of the hierarchical sweeping module exacerbates this phenomenon.
Fig. \ref{h}(d) verifies the decimeter-level UE localization ability ($<0.4\;{\rm m}$) of ``Proposed Method’’ in changeable wireless environments. 
The experiments have similar performance at $t \in {(1,20) }{\;\rm s}$. Then, the angle estimation error from the hierarchical sweeping module affects ``Hierarchical Method’’ at $t \in {(20,40) }{\;\rm s}$, where ``Proposed Method’’ outperforms it by ${{\rm{E}}_{{\rm{UE}}}}$ reduction of $0.25{\;\rm m}$.
Fig. \ref{h}(e) verifies the decimeter-level radio map construction ability ($<0.25\;{\rm m}$) of ``Proposed Method’’ in changeable wireless environments. 
There exists a sudden increase of OSPA at $t=10{\;\rm s}$. The reason is that the corresponding emerging LoS path from BS is close to the presenting NLoS path from VA${\rm _2}$ in the aspect of AoA. 
Their performance get stable at $t \in {(11,30) }{\;\rm s}$, where ``Proposed Method’’ outperforms ``Hierarchical Method’’ by OSPA reduction of $0.1{\;\rm m}$. Then, OSPA increases at $t \in {(30,40) }{\;\rm s}$ because of the path number reduction.

{\color{black}
\subsubsection{Stable Wireless Environment}
Scenario 2 in Fig. \ref{g}(b) depicts the stable wireless environment where no birth or death of the channel path occurs. In this scenario, the UE remains in the area where one LoS path from the BS and three NLoS paths from VA$_1$-VA$_3$ are present.
We assume that the UE trajectory between time $t$ and $t-1$ satisfies $ {\left\| {{\mathbf{x}^{(t )}} - \mathbf{x}^{(t-1 )}} \right\|}_2\sim \mathcal {U} \left({0.7{\;\rm m},0.9{\;\rm m}} \right)$. 
We then generate random UE trajectories consisting of $30$ steps to perform 5000 Monte-Carlo simulations.
In addition, we assume that the experiments know the location of the BS and initial UE position, and that the hierarchical sweeping module is performed during the first three steps for initialization.

\begin{table*}[]
\caption{\color{black}Parameters for Beam Management Simulations}\vspace{-6mm}
\label{table3}
\center
\begin{tabular}{|l|l|l|l|}
\hline
Path Number & $L=3\;\left(1\right)$ & \multirow{2}{*}{Path Gain} & ${g_0}\sim \mathcal {U} \left({0.7,0.9} \right)$ \\ \cline{1-2} \cline{4-4} 
Codebook Resolution & $N=512\;\left(256\right)$ & & ${g_l} \sim \mathcal {U} \left({0.3,0.5} \right),\;l\ge1$ \\ \hline
\multirow{2}{*}{Antenna Number} & ${N_{{\rm{BS}}}} = 32,{\rm{\;}}{N_{{\rm{UE}}}} = 16$ & \multirow{3}{*}{Prior Angle Information} & $\theta _{{\rm{scale}},l}^{},\phi _{{\rm{scale}},l}^{} = 0.1\pi\; {\rm{rad}}$ \\ \cline{4-4} 
& $ \left({N_{{\rm{BS}}}} = 16,{\rm{\;}}{N_{{\rm{UE}}}} = 8 \right) $ & & \multirow{2}{*}{\begin{tabular}[c]{@{}l@{}}${\theta _{{\rm{VA}},l}} \sim \mathcal {U} \left({{\theta _l} - 0.25{\theta _{{\rm{scale}},l}},{\theta _l} + 0.25{\theta _{{\rm{scale}},l}}} \right)$\\ ${\phi _{{\rm{VA}},l}}\sim \mathcal {U} \left({{\phi _l} - 0.25{\phi _{{\rm{scale}},l}},{\phi _l} + 0.25{\phi _{{\rm{scale}},l}}} \right)$\end{tabular}} \\ \cline{1-2}
Path Angle & ${\theta _l},{\phi _l} \sim \mathcal {U} \left({0,\pi } \right) \;{\rm rad}$ & & \\ \hline
\end{tabular}
\vspace{-8mm}
\end{table*}

Fig. \ref{i} shows the results of ``Proposed Method’’, ``Hierarchical Method’’, and ``Without IMU’’ in Scenario 1. 
In this figure, the red curves represent ``Hierarchical Method’’, the blue ones represent ``Proposed Method’’, and the yellow ones represent ``Without IMU’’. 
The gray bar describes the change of path number $L^{(t)}$. The analysis of the figure is as follows. 

Fig. \ref{i}(a)-(c) illustrate that ``Proposed Method’’ can realize better angle estimation and get higher SE with around $44\%$ less overhead consumption compared with ``Hierarchical Method’’ and ``Without IMU’’ in stable wireless environments. 
Additionally, Fig. \ref{i}(a)-(c) show that ``Hierarchical Method’’ and ``Without IMU’’ approaches exhibit similar performance. 
Fig. \ref{i}(d) verifies the centimeter-level UE localization ability (around $0.05\;{\rm m}$) of ``Proposed Method’’ in stable wireless environments. 
``Proposed Method’’ slightly outperforms ``Hierarchical Method’’ and exhibits an error reduction of around $0.2\;{\rm m}$ compared to ``Without IMU’’. 
Fig. \ref{i}(e) verifies the decimeter-level radio map construction ability (around $0.25\;{\rm m}$) of ``Proposed Method’’ in stable wireless environments. 
The radio map constructed by ``Proposed Method’’ has similar accuracy to ``Hierarchical Method’’ and has an OSPA reduction of around $0.4\;{\rm m}$ compared to ``Without IMU’’.
}

To summarize, radio map construction (with OSPA $<0.5 \;{\rm m}$) and UE localization (with an error $<0.4 \;{\rm m}$) can be realized by the proposed joint design in changeable wireless environments. 
High-precision radio map construction (with OSPA around $ 0.25\; {\rm m}$) and centimeter-level UE localization (with an error around $0.05 \;{\rm m}$) can be realized in stable wireless environments. 
The introduction of IMU can improve UE localization (with an error reduction around $0.3\;{\rm m}$) and radio map construction accuracy (with an OSPA reduction $0.4\;{\rm m}$), particularly in changeable wireless environments. 
The introduction of the switching module can enable the feature-aided tracking module (with an overhead reduction of around $ 40\%$) and improve the angle estimation accuracy, particularly in stable wireless environments.

\subsection{Performance of Proposed Modules}
\vspace{-1.5mm}
In Section IV-A, ``Proposed Method’’ can outperform ``Hierarchical Method’’ when the feature-aided tracking module with a reliable radio map is activated. 
To evaluate the hierarchical sweeping module, the feature-aided tracking module, and the switching module thoroughly, a set of 5000 Monte-Carlo simulations are conducted. 
TABLE \ref{table3} presents the parameters. The path number, codebook resolution, and antenna number are set as $L=3$, $N=512$, and $N_{\rm BS}=32,N_{\rm UE}=16$, respectively.

\begin{figure*}
\centering
\subfloat[]{
\hspace{-4.5mm}\includegraphics[scale=0.46]{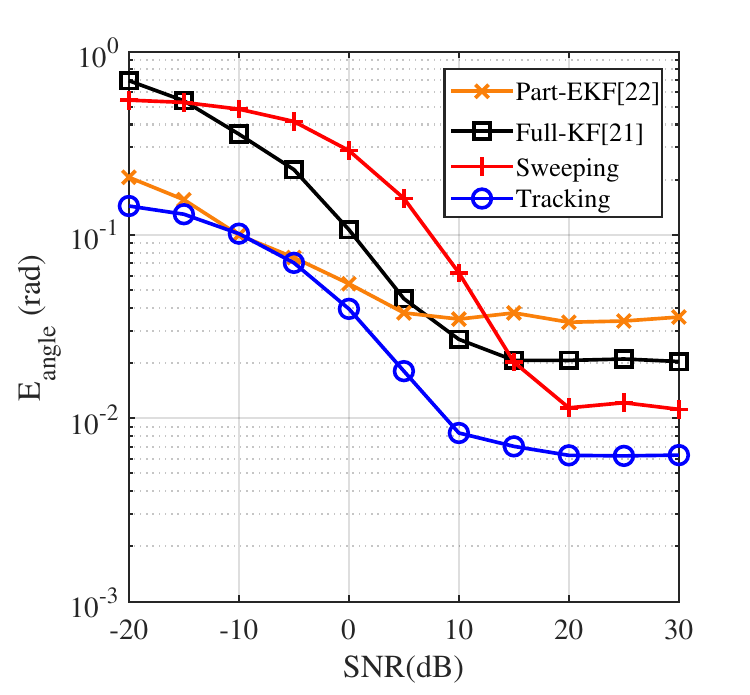}\hspace{-5mm}
}
\quad
\subfloat[]{
\hspace{-6mm}\includegraphics[scale=0.46]{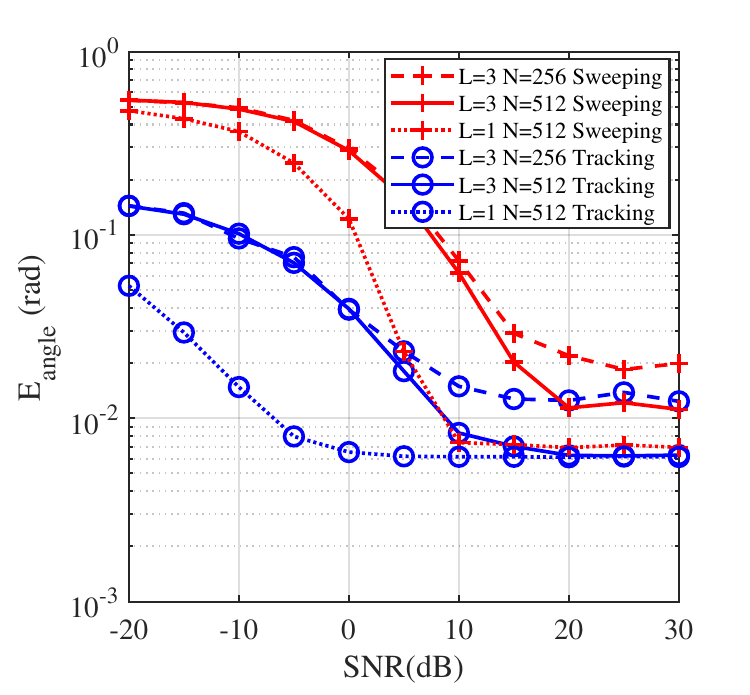}\hspace{-7mm}
}
\quad
\subfloat[]{
\hspace{-4.5mm}\includegraphics[scale=0.46]{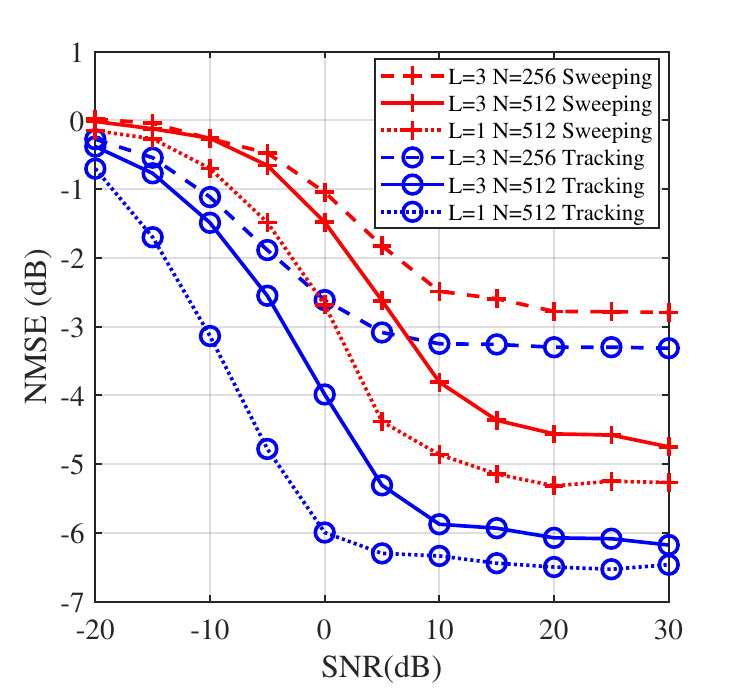}\hspace{-5mm}
}
\quad
\subfloat[]{
\hspace{-0mm}\includegraphics[scale=0.46]{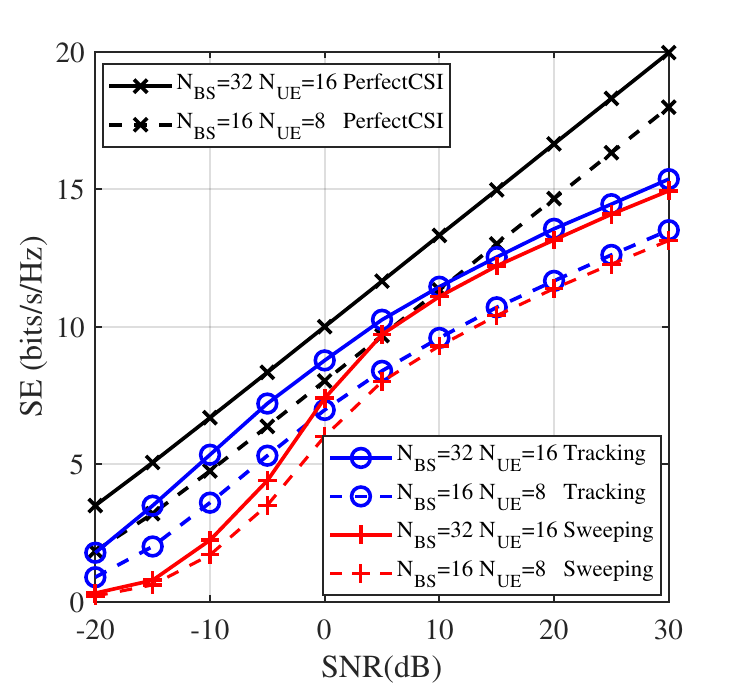}\hspace{-0mm}
}
\quad
\subfloat[]{
\hspace{-0mm}\includegraphics[scale=0.46]{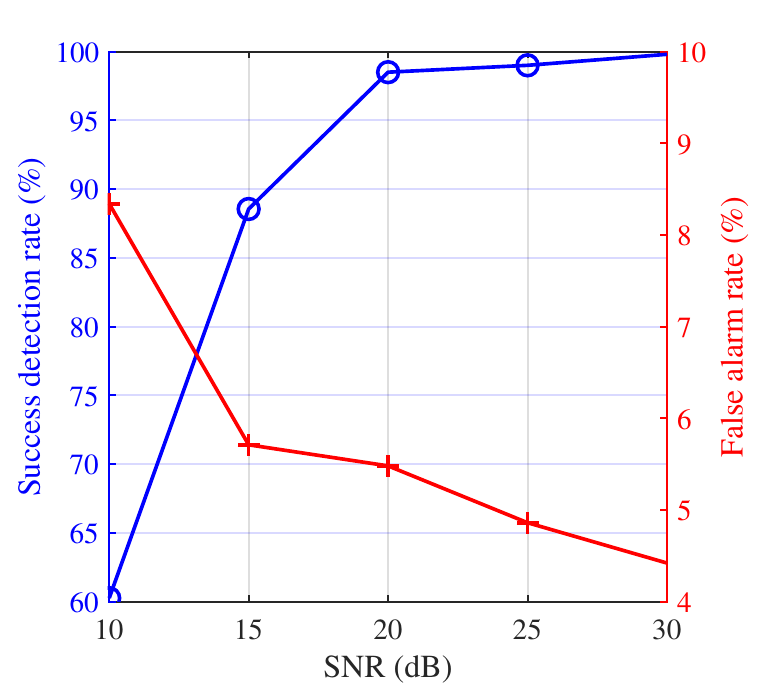}\hspace{-0mm}
}
\caption{(a) Angle estimation performance of different schemes; (b) Angle estimation performance, (c) channel estimation performance, and (d) SE performance of the hierarchical sweeping and feature-aided tracking modules with different path number $L$, codebook resolution $N$, and antenna number $N_{\rm BS},N_{\rm UE}$; (e) Success detection rate and false alarm rate of the switching module. }
\label{f}
\vspace{-10mm}
\end{figure*}

{\color{black}
\subsubsection{Benchmark Comparison}
In this subsection, \cite{a1001,a100101} based on Kalman filters are set as benchmarks for the proposed modules, called Full-KF and Part-EKF, respectively. 
Full-KF is implemented with $M_{\rm BS}M_{\rm UE}$ measurements covering fixed angle domain for angle estimation, where we set $M_{\rm BS}=32$ and $M_{\rm UE}=16$. 
Part-EKF is implemented with $4L$ measurements pointing to a specific angle provided by Kalman prediction. 
These benchmarks assume that the AoA and AoD change slowly and that the path number is available. 
We evaluate the angle estimation accuracy, time complexity, and computational complexity for comparison based on 5000 Monte Carlo simulations using the parameters in Table \ref{table3}. 

The angle estimation accuracy of different schemes for varied SNRs is shown in Fig. \ref{f}(a), where ``Sweeping'' and ``Tracking'' curves represent the hierarchical sweeping module and the feature-aided tracking module, respectively. 
As expected, all the schemes achieve better accuracy as SNR increases. 
The feature-aided tracking module outperforms the others over the SNR, while the others have different advantages in different SNR regions. 
In the low SNR region, Part-EKF and the feature-aided tracking module covering a limited angle domain can reduce the effect of noise and have better performance over Full-KF and the hierarchical sweeping module. 
In the high SNR region, the performance of Part-EKF is limited by inadequate beam measurements. 
Additionally, Full-KF and the hierarchical sweeping module are affected by multipath interference. Therefore, the feature-aided tracking module can outperform these three schemes.

The time and computational complexities are discussed in TABLE IV. 
The time complexities are calculated based on the number of the beam measurements required for one single estimation, where the exact values are given based on the parameters in TABLE \ref{table3}.
Full-EKF and Part-EKF have the highest and lowest time complexity, respectively. 
The feature-aided tracking module has a time complexity that is around $33.4\%$ lower than that of the hierarchical sweeping module. 
{\color{black}
The computational complexities are calculated based on the variables of codebook resolution $N$ and channel path number $L$, taking into account both codebook generation and beam management procedures.\footnote{{\color{black}{The codebooks of the hierarchical sweeping module and Full-EKF can be partially pre-generated to trade off storage complexity for reduced computational complexity, thus accommodating different hardware conditions.}}} Among the methods considered, Full-EKF has the highest computational complexity, while Part-EKF has the lowest, which is solely dependent on the path number $L$. The feature-aided tracking module is capable of reducing computational complexity compared to the hierarchical sweeping module by leveraging prior angle information.
}

To summarize, the feature-aided tracking module can get accurate angle estimation with relatively low time and computational complexities compared to other benchmarks. 
However, the beamforming vectors of the feature-aided tracking module are designed to point at arbitrary angles, and the prior angle information relies on results from the sensing services that require IMU and SLAM.
}
\begin{table*}[]
\caption{\color{black}Algorithm Complexity Comparison}\vspace{-12mm}
\label{table4}
\center
\begin{tabular}{|c|ll|l|c|}
\hline
Algorithm & \multicolumn{2}{c|}{Time Complexity} & \multicolumn{1}{c|}{Computational Complexity} & Extra Demand \\ \hline
Sweeping & \multicolumn{1}{l|}{$\left(4{{\rm{log}}_{2}N}\right)L$} & $108$ & $\mathrm{O}\left(L N \log _{2} N+L \log _{2} N+L\right) $ & None \\ \hline
Tracking & \multicolumn{1}{l|}{ $\mathop \sum \nolimits_{l = 0}^{L-1} 4 \left({{{\rm{log}}_{2}N} - {j_l}} \right) $ } & $72$ & $\mathrm{O}\left(LN \log _{2}\left( \frac{{\phi_{\rm {scale}}}}{\pi}N\right)+L\log _{2}\left(\frac{ {\phi_{\rm {scale}}}}{\pi}N\right)+L\right) $ & Prior angle information \\ \hline
Full-KF & \multicolumn{1}{l|}{$M_{\rm BS}M_{\rm UE}$} & $1024$ & ${\mathrm{O}}\left(\left(M_{\mathrm{UE}} M_{\mathrm{BS}}\right)^{3}+L\left(M_{\mathrm{UE}} M_{\mathrm{BS}}\right)^{2}+L^{2} M_{\mathrm{UE}} M_{\mathrm{BS}}+L^{3}\right)$ & \multirow{2}{*}{\begin{tabular}[c]{@{}c@{}}Angle changes slowly;\\ Path number available\end{tabular}} \\ \cline{1-4}
Part-EKF & \multicolumn{1}{l|}{$4L$} & $12$ & ${\mathrm{O}(L)}$ & \\ \hline
\end{tabular}\vspace{-8mm}
\end{table*}
%Compare the computational complexity. Calculate the computational complexity when 

\subsubsection{Impact of path number, code resolution, and number of antennas}
The parameters in TABLE \ref{table3} are set with different values ($L=1$, $N=256$, or $N_{\rm BS}=16,N_{\rm UE}=8$) separately for performance evaluation of the hierarchical sweeping and the feature-aided tracking modules. 
%We assume that reliable prior angle information is available for the feature-aided tracking module in this section. 
The channel path number $L$ is perfectly known in this subsection because the path number estimation error of the hierarchical sweeping module can bring unfairness. 
The performance metrics are set according to Section IV: ${{\rm{E}}_{{\rm{angle}}}}$ in (\ref{29}) with $L_{\Delta}=0$ is set for evaluating sensing ability. Then, NMSE in (\ref{32}) and ${\rm{S}}{{\rm{E}}_{{\rm{IP}}}}$ in (\ref{40}) are set for evaluating communication performance. 
Fig. \ref{f} depicts the performance curves SNR-${{\rm{E}}_{{\rm{angle}}}}$/NMSE/SE. 
The red and blue curves represent hierarchical sweeping and feature-aided tracking modules, respectively. The performance analysis is as follows.

The sensing ability is evaluated by SNR-${{\rm{E}}_{{\rm{angle}}}}$ curves shown in Fig. \ref{f}(b), where the solid curve represents the original experiment with $L=3,N=512$, the scatter curve represents the path number control experiment with $L=1,N=512$, and the dotted curve represents the codebook resolution control experiment with $L=3,N=256$. 
{\color{black}First, ${{\rm{E}}_{{\rm{angle}}}}$ decreases as SNR increases, and it reaches the lower bound (varying from $2\times 10^{-2}\;{\rm rad}$ to $6\times 10^{-3}\;{\rm rad}$) when SNR is approximately $15\;{\rm dB}$, where the lower bound is limited by the angular resolution ${\rm{\pi }}/N$.}
Then, the feature-aided tracking module outperforms the hierarchical sweeping module over the SNR region because the prior angle information can reduce the angle searching range in the feature-aided tracking module. 
Finally, ${{\rm{E}}_{{\rm{angle}}}}$ is negatively correlated with the codebook resolution $N$ and positively correlated with the path number $L$ in the high SNR region. 
The reason is that ${{\rm{E}}_{{\rm{angle}}}}$ is lower bounded by the codebook resolution ${\pi }/{{N}}\;{\rm rad}$, and the modules pass errors of the already estimated paths through residual channel calculation in (\ref{9}). 
In addition, the hierarchical sweeping module suffers more performance reduction with the increase of path number $L$ compared with the feature-aided tracking module, as shown in Fig. \ref{f}(b). 
The prior angle information enables the feature-aided tracking module to isolate the noise and possible path interference beyond the angle searching range, leading to higher angle estimation accuracy compared with the hierarchical sweeping module.

The communication performance is evaluated by SNR-NMSE/SE curves, as shown in Fig. \ref{f}(c) and (d). The representations of curves in Fig. \ref{f}(c) remain the same as those in Fig. \ref{f}(b). 
Similar conclusions can be achieved that the feature-aided tracking module outperforms the hierarchical sweeping module in the aspect of channel estimation over the SNR region, and the channel estimation accuracy is positively correlated with $N$, and negatively correlated with $L$ in high SNR region. 
The difference between curves in Fig. \ref{f}(c) in high SNR region is more evident than those in Fig. \ref{f}(b). The path gain estimation error is correlated with the angle estimation error. 
Therefore, channel estimation based on the path gain and angle estimation is more sensitive. 
Fig. \ref{f}(d) shows the SE performance, where different antenna number settings are considered. SE with perfect CSI in (\ref{34}) is drawn in black as the upper bound.
The feature-aided tracking module gets higher SE compared with the hierarchical sweeping module over the SNR region, and the SE is positively correlated with antenna numbers $N_{\rm BS},N_{\rm UE}$. 
The reason is that antenna arrays with more antenna elements generate beams with narrower widths, leading to less energy loss.

To conclude, the feature-aided tracking modules can complete high-accuracy angle estimation with relatively low time and computational complexities, compared with the hierarchical sweeping module and schemes based on Kalman filters \cite{a1001,a100101}. 
The feature-aided tracking module outperforms the hierarchical sweeping module with reduced overhead by approximately $40\%$.

{\color{black}
\subsubsection{Performance of Switching Module}
To evaluate the switching module, two metrics have been proposed: success detection rate and false alarm rate. 
The former measures the percentage of correctly switching between the hierarchical sweeping module and the feature-aided tracking module when the path number changes, while the latter measures the false switch when the path number remains unchanged. 
The effectiveness of the switching module is evaluated using these two metrics with 5000 Monte-Carlo simulations for different SNRs. 
The success detection rate and false alarm rate curves are shown in Fig. \ref{f}(e).

As seen from Fig. \ref{f}(e), with the increase of SNR, the change trends of the success detection rate and false alarm rate are opposite. 
These two metrics depend on the same thresholds $E_{\rm{min}}$ and $\Delta _E^{\rm min} $ in (\ref{28}) controlling the change of switching symbol ${D^{(t)}}$. 
Therefore, there is a tradeoff between the success detection rate and the false alarm rate. 
A high success detection rate can detect path number change accurately, while a low false alarm rate can effectively reduce overhead consumption. 
For SNRs greater than $15\;{\rm{dB}}$, the joint design maintains the success detection rate above $80\%$. As the SNR increases, the false alarm rate reduces from $6\%$ to $4\%$.
These results indicate that the proposed switching module achieves a good balance between reducing overhead and detecting path number changes.

}
\vspace{-3.5mm}
% Conclusion
\section{Conclusion}
\vspace{-1.5mm}
This study investigated a joint beam management and SLAM design. 
The beam management module estimates channel parameters, while SLAM utilizes the angle estimates with the help of IMU to perform UE localization and radio map construction. 
The switching module allows the design to adapt to different wireless environments. Performance metrics were introduced for sensing and communication. 
The simulation results demonstrate that the proposed joint design can achieve overhead-reduced beam management with a reduction of around $40\%$, high-accuracy radio map construction with an error of less than $0.5 \;{\rm m}$, and UE localization with an error of less than $0.4 \;{\rm m}$ in different wireless environments.

\bibliographystyle{IEEEtran}
\bibliography{ bmlong}

\end{document}